\documentclass[reviewcopy,12pt]{elsarticle}

\usepackage[reviewcopy]{adndt}
\usepackage{longtable}

\usepackage{amsmath}

\usepackage{amssymb}

\biboptions{square,sort&compress}
\bibpunct[]{[}{]}{,}{n}{}{;}
\usepackage{hyperref} 

\begin{document}

\begin{frontmatter}

\journal{Atomic Data and Nuclear Data Tables}


\title{Table of electronic factors for E0 electron and electron-positron pair conversion transitions}

  \author[ANU]{J.T.H.~Dowie}
  \author[ANU]{T.~Kib\'edi\corref{cor1}}
  \ead{Tibor.Kibedi@anu.edu.au}
  \author[ANU]{T.K.~Eriksen \fnref{fn1}}
  \author[ANU]{A.E.~Stuchbery}
  \cortext[cor1]{Corresponding author.}
  \fntext[fn1]{Present Address: Department of Physics, University of Oslo, N-0316 Oslo, Norway}

  \address[ANU]{Department of Nuclear Physics, Research School of Physics and Engineering,
                The Australian National University, Canberra, ACT 2601, Australia}

\date{\today} 

\begin{abstract}
 A new tabulation of electronic factors is reported for electron conversion for elements of Z
 from 5 to 126 and electronic factors for electron-positron pair conversion for elements of even
 Z from 4 to 100.
 The electronic factors for electron conversion, $\Omega_{CE}($E0$)$, were calculated using a
 modified version of the \emph{CATAR} program developed by Pauli and Raff with a
 relativistic-Hartree-Fock-Slater approach (Comp. Phys. Comm. \textbf{9} (1975) 392).
 The electronic factors for electron-positron pair conversion, $\Omega_{IPF}($E0$)$, were
 calculated using the model developed by Wilkinson (Nucl. Phys.  \textbf{A133} (1969) 1).
 The data tables presented here cover all atomic shells up to R2 and transition energies
 from 1 keV to 6000 keV and from 1100 keV to 8000 keV for pair conversion.
 A comparison with previous electronic factor tabulations is presented.
 Ratios of experimental $\Omega($E0$)$ values for 83 E0 transitions in $8 \leq ~Z~ \leq 98$
 are compared to this tabulation.
 Two examples of how to use the tabulation to extract E0 strengths are also included.

\end{abstract}

\end{frontmatter}




\newpage

\tableofcontents
\listofDtables
\listofDfigures
\vskip5pc

\section{Introduction}
\label{sec:intro}

 Electric monopole (E0) transitions connect quantum states of the same spin-parity
 in atomic nuclei and are associated with collective excitations and phenomena such as
 shape coexistence, isotope and isomer shift, and volume oscillations.
 After more than 80 years since the first suggestion of an E0 transition,
 the 1.426 MeV transition in Radium C ($^{214}$Po) by
 Ellis and Aston \cite{i1930Ellis}, the characterization of E0
 transitions still remains a challenge for both experimental nuclear
 spectroscopy and nuclear theory \cite{i1999Wo07,i2005Ki02,i2011He11,i2012Ze02}.

 The geometrical model of Bohr and Mottelson \cite{i1975Bohr} predicts two
 different types of oscillations in deformed nuclei with an ellipsoidal shape:
 \emph{(a)} the $\beta$-vibration which preserves axial symmetry,
 and \emph{(b)} the $\gamma$-vibration which does not.
 Rotational structures built on these  states can be observed in the low energy spectrum of many
 deformed even-even nuclei.
 Transitions from the $0^{+}$ $\beta$-vibrational state to the ground state were considered as the classical
 examples of pure E0 transitions.
 In nuclei of spherical ground state, two-phonon excitations also can produce excited  $0^{+}$ states.
 However, recently, both of these vibrational models have been questioned \cite{2012Wood_JPCS}.
 An alternative approach, developed initially for the interpretation of $0^{+}$ states in the vicinity of
 double-magic nuclei, has been proposed.
 Classical examples of these multiparticle multihole excitations across the shell gap
 are the $0^{+}$ states in $^{16}$O
 and $^{40}$Ca, which in fact are the first-excited states in both nuclei.
 These states are deformed and in single closed-shell nuclei bands built on such excited $0^{+}$ states 
 have been observed in many cases.
 A distinct feature of these nuclei is the co-existence of spherical and deformed states
 at similar excitation energies and  the presence of E0 transitions.

 While in some cases it is possible to develop a consistent description of electric
 monopole transitions \cite{i2012Ze02}, further theoretical progress is required for a unified picture.
 Our experimental knowledge of pure E0 transitions, those between $0^{+}$ states,
 including branching ratios and absolute transition rates, B(E0), is far from complete.
 Pure E0 transitions are expected over the entire nuclear landscape.
 For a recent review, see  \cite{i2005Ki02}.
 For $J \rightarrow J$ transitions, where $J > 0$, E0 transitions mix with E2 and M1
 transitions giving, in the general case, transitions of mixed E0+E2+M1 multipolarity.
 The last review of these type of electromagnetic transitions, by Lange, Kumar
 and Hamilton \cite{i1982La26} included the E0/E2 and E2/M1 mixing ratios, which are
 required to fully characterize such transitions. However, this review was published more than 30 years ago.
 A more recent compilation of mixed E0+E2+M1 transitions can be found in Wood \textit{et al.} \cite{i1999Wo07},
  but a full review of the subject is long overdue.

 Central to the characterization of the E0 transitions is the knowledge of the E0 electronic factors, $\Omega(E0)$.
 The E0 transition rate can be factorized into an electronic and a nuclear contribution.
 The nuclear component contains all of the information about the nuclear structure while
 the electronic factor represents the contribution of the atomic electrons to the transition rate.
 The E0 electronic factors cannot be measured directly but can be calculated theoretically.
 Without the electronic factors, the nuclear component cannot be extracted from the E0
 transition rate.

\emph{BrIcc} is a currently existing database of the most up-to-date and most accurate internal conversion and
 pair-conversion coefficients for elements of $Z$ between 5 to 126 \cite{i2008Ki07}.
 The conversion coefficients are obtained from the Dirac-Fock calculations of Band \textit{et al.}
 \cite{i2002Ba85}, and the pair-conversion coefficients from the work of Soff and Hofmann
 \cite{i1981So10, i1996Ho15}.
 The current $\Omega($E0$)$ table adopted for \emph{BrIcc} \cite{i2008Ki07} is based on three
 different calculations \cite{i1969Ha61,i1970Be87,i1986PaZM}, which have been calculated
 using different physical assumptions, and cover only the K, L1 and L2 atomic shells; the energy and $Z$ range
 is also limited.
 $\Omega($E0$)$ pair conversion coefficients were calculated only for $8 \leq Z \leq 40$ atomic numbers.
 Therefore, a new, more complete tabulation is warranted.

 In this paper, we present tables of electronic factors, $\Omega($E0$)$ of electric
 monopole conversion probabilities for all atomic shells up to the R2  with transition
 energies from 1 keV above atomic shell binding energies to 6000 keV.
 New tabulations of $\Omega_{IPF}($E0$)$  for pair conversion are also presented.
 These tables cover even $Z$ in the range of 4 to 100 and transition energies from 1100 keV to 8000 keV.
 This coverage is compatible with the range of the recent electron conversion coefficient calculations by
 Kib\'edi \textit{et al.} \cite{i2008Ki07}.
 The numerical calculations have been carried out with a modified version of the \emph{CATAR} program
 by Pauli and Raff \cite{i1975Pa26} and using the self-consistent-field calculated via a modified
  version of the \emph{HEX} code by Liberman, Cromer and Waber \cite{i1971Liberman}.
 The details of the programs used and the modifications made will be discussed in \ref{sec:Calculations_CE}.
 The $\Omega_{IPF}($E0$)$ values were calculated using \emph{WspOmega}, a new code based on the Wilkinson
 formulation \cite{i1969Wi29}.

\section{E0 transitions and $\Omega($E0$)$ electronic factors}
\label{sec:theory}

 The theoretical background for the emission of radiation via E0 transitions is
 well documented \cite{i1956Ch21,i1970Be87,i1975Pauli,i1987Kr19,i2005Ki02}.
The electric monopole operator couples the nucleons to the atomic electrons and the Dirac
 sea.
 This interaction is completely localized within the nucleus and therefore the E0 transition
 rate depends strongly on the charge distribution of the nuclear quantum states.
 It is widely accepted that E0 transitions depend on certain details
 of the nuclear structure and can be used as a sensitive probe of nuclear models.

 Electric monopole transitions proceed via the emission of a conversion electron
 or, for nuclear transitions of energy greater
 than twice the electron rest mass, through the emission of an electron-position pair.
 The emission of a single $\gamma$ ray is strictly forbidden due to the conservation
 of the angular momentum, and the probability of emission of two $\gamma$ rays or de-excitation via
 any other two-quantum process is sufficiently low as to be negligible \cite{i1987Kr19}.
 Two-quantum processes are not considered in this work.
 With these assumptions, the E0 transition probability can be expressed as
 \begin{equation}
 \label{eqn:WE0}
 W(\text{E0}) = \frac{1}{\tau (\text{E0})} = W_{ce}(\text{E0}) + W_{IPF}(\text{E0}) \, ,
 \end{equation}
 where $\tau(\text{E0})$ is the partial mean life of the excited state with respect to the
 E0 decay, and $W_{ce}$ and $W_{IPF}$ are the transition probabilities for internal conversion
 and pair emission,  respectively.

 Church and Weneser \cite{i1956Ch21} were the first to introduce the so-called electronic
 factor, $\Omega ($E0$)$, to evaluate the E0 transition probability, $W($E0$)$.
 The E0 transition probability can be separated into an atomic ($\Omega($E0$)$) and a nuclear
 ($\rho($E0$)$) component:
 \begin{equation}
   \label{eqn:omegarho}
   W(\text{E0}) = \Omega(\text{E0}) \times |\rho(\text{E0})|^2 \, .
 \end{equation}
 The dimensionless monopole strength parameter, $\rho($E0$)$,  contains all of the information about
 nuclear structure and is related to the monopole matrix element, $M($E0$)$, by
 \begin{equation}
   \label{eqn:rhoMe0}
   \rho(\text{E0}) = \frac{\langle f | M(\text{E0}) | i \rangle}{eR^{2}} \, ,
 \end{equation}
 where $e$ is the electronic charge, $R = R_{\circ} A^{1/3}$ is the nuclear radius, $A$ is the atomic
 mass number, and $R_{\circ} = 1.20$ fm.
 The electronic factor is largely independent of the nuclear properties, but it depends
 on the atomic number, the nuclear transition energy, and in the case of electron
 conversion, the atomic shell involved.

\subsection{Internal Conversion and $\Omega($E0$)$ electronic factors}
\label{sec:ICandOmega}
 The internal conversion pathway of the E0 transition is well studied and the
 details of the formulae and assumptions used here are described in detail in the
 work of Pauli, Alder, and Steffen \cite{i1975Pauli}.
 The E0 transition rate can be expressed as
 \begin{equation}
   \label{eqn:WE0.CE}
  W_{ce}(\text{E0}) = 2\pi(j+1)  \frac{|\langle I \kappa || H^{E}_{0} || I \kappa \rangle|^2}
         {2 I+1} \, ,
 \end{equation}
 where $j$ and $\kappa$ are the total and relativistic angular momentum of the
 emitted electron, and $I$ is the initial and final nuclear spin, and
 $|\langle I \kappa || H^{E}_{0} || I \kappa \rangle|$
 is the reduced E0 transition matrix element.

 The E0 transition matrix element for internal conversion is
 \begin{equation}
   \label{eqn:E0MatEl}
   \langle f | H_{00} | i \rangle = \int_{0}^{\infty} dV_n \rho_n (r_n) \int_{0}^{r_n} dV_e
   \rho_e (r_e) \Bigg( \frac{1}{r_n} - \frac{1}{r_e} \Bigg) \, ,
 \end{equation}
  where the subscripts $n$ and $e$ denote nuclear and electron coordinates, respectively.
 The electron density, $\rho_e (\mathbf{r_e})$, can be replaced by the explicit expression
 in terms of the radial Dirac electron functions, giving:
 \begin{equation}
  \label{eqn:E0MatElradial}
  \langle f | H_{00} | i \rangle = -e \int_{0}^{\infty} dV_n \rho_n (r_n) \int_{0}^{r_n}
  dr_e [u_{\kappa}(r_e) u_{\kappa_{\circ}} (r_e) + v_{\kappa}(r_e) v_{\kappa_{\circ}}(r_e)]
  \Bigg( \frac{1}{r_n} - \frac{1}{r_e} \Bigg) \, ,
 \end{equation}
  where $\kappa_{\circ}$ and $\kappa$ are the Dirac quantum numbers for the bound and
 continuum conversion electron, respectively.
 The $u_{\kappa}(r_e)$ and $v_{\kappa}(r_e)$ are the large and small relativistic radial
 functions for the electron.
 They are defined through the relativistic electron wave function:
 \begin{equation}
  \label{eqn:diracDef}
    \langle \textbf{r} | \kappa \mu \rangle = \psi_{\kappa \mu}(r_{~}) = \frac{1}{r}
    \begin{pmatrix}  u_{\kappa}(r) &  \phi_{\kappa \mu}(\hat{r}) \\ iv_{\kappa}(r) &
    \phi_{-\kappa \mu}(\hat{r}) \end{pmatrix} \, ,
 \end{equation}
 where $\phi_{\kappa \mu}$ are the Dirac angular momentum spinors.

 Inside the nucleus, the radial functions can be expanded into a power series
 (see Eqn. 10.180 in \cite{i1975Pauli}).
 Substituting the power series for the radial functions in Eqn.~(\ref{eqn:E0MatElradial}),
 and integrating over the electron coordinates, we obtain
\begin{equation}
  \label{eqn:Mat}
   \langle f | H_{00} | i \rangle =
   \int_{0}^{\infty} dV_n \rho_n(\textbf{r}_n) ( \sum_{m=0}^{\infty}
   \frac{\gamma_m}{(\bar{p}+2m)(\bar{p}+2m+1)} (\frac{r_n}{R})^{2m+\bar{p}}) \, ,
 \end{equation}
 where the $\gamma_m$ are the power series parameters for the radial wave functions,
  and $\bar{p}$, for E0 transitions, is $2|\kappa|$.
 From here, we can apply the Wigner-Eckhart theorem, by extracting $\Omega($E0$)$, and obtain $\rho($E0$)$.

 The monopole transition strength can be expressed \cite{i1975Pauli} as
 \begin{equation}
 \label{eqn:RhoE0}
   \rho(\text{E0}) =
   \int d V_{n} \rho_{n}(r_{n})
   \left( \frac{r_{n}}{R} \right)^{2 | \kappa_{\circ} |}
   \left[ 1 + \left( \frac{r_{n}}{R} \right)^{2} \frac{\gamma_{1}}{\gamma_{\circ}}
   \frac{2 | \kappa_{\circ} | (2 | \kappa_{\circ} | +1)}
       {(2 | \kappa_{\circ} | + 1 )(2 | \kappa_{\circ} | +3)} + ... \right] \, ,
  \end{equation}
 and the $\Omega_{ce}(\kappa_{\circ})$ electronic factor is defined as
 \begin{equation}
   \label{eqn:Omg}
   \Omega_{ce}(\kappa_{\circ}) =
   2 \pi (2j+1) \left( \frac{e \gamma_{\circ} } {2 | \kappa_{\circ} | (2| \kappa_{\circ} | +1)} \right) ^{2} \, .
 \end{equation}

 The $\Omega_{ce}$ electronic factor is related to the so-called E0 coefficient given by
  Hager and Seltzer \cite{i1969Ha61}, $A_{ce}($E0$)$, by
 \begin{equation}
   \Omega_{ce} = 8 \pi \alpha k \times A_{ce}(\text{E0}) \times C \, ,
  \end{equation}
 where $\alpha$ is the fine structure constant, $k$ is the transition energy in $m_0 c^2$ units,
  and $C = 7.7631 \times 10^{20}$ is a unit conversion factor.

\subsection{Internal pair formation and $\Omega($E0$)$ electronic factors}
\label{sec:IPFandOmega}

 For transitions with energies larger than twice the electron rest mass, an alternative process
 involving the emission of electron-positron pairs can compete with the conversion electron process.
 The theoretical development for internal pair formation is less well studied and a brief history of the
 development will be given here.
 The first theoretical description of internal pair conversion was given in 1933 by Nedelsky and
  Oppenheimer \cite{i1933Nedelsky} for transitions of angular momentum $L>0$ and the presence of
  electron-positron pairs for E0 transitions was first suggested by Fowler and Lauritsen
  \cite{i1939Fowler} in 1939.
  Thomas \cite{i1940Thomas} developed a more detailed model after the first calculation of the pair
  emission rate by Oppenheimer and Schwinger \cite{i1939Oppenheimer}.
 Using the Born approximation, Oppenheimer in 1941 \cite{i1941Oppenheimer} derived a simple equation
 for the energy and angular correlation of E0 pair emission.
 These expressions had limited validity for higher $Z$ systems and for lower energy transitions.
 They did not take into account the finite size of the nucleus, the effect of atomic screening, and
  the nuclear Coulomb effects.


 The next development in the description of internal pair formation for E0 transitions was described by
 Wilkinson in 1969 \cite{i1969Wi29}.
 This formulation takes into account the effect of the nuclear Coulomb field on the emission process,
 the Coulomb correction, and the effect of atomic screening is also corrected for using a
 Thomas-Fermi-Dirac statistical model of the atom following the works of Durand, and
 Bahcall~\cite{i1964Durand, i1966Bahcall}.

 The most realistic model for internal pair formation was developed in 1981 by Soff \cite{i1981So10}.
 This model takes into account the finite size of the nucleus. Later in the 1990s Hofmann and Soff
 \cite{i1996Ho15} extended the model to evaluate the double differential emission probability,
 $d\Omega_{IPF}($E0$)/dE\,d \Theta_{\pi}$, where $\Theta_{\pi}$ is the separation angle between the
 electron and the positron.
 The theory of internal pair formation is well described in the work of Hofmann \textit{et al.}
 \cite{i1990Hofmann} and the key points will be summarized here.

 Internal pair formation results in the emission of an electron-positron pair with the kinetic energies of
 the electron and positron summing to that of the transition energy minus two times the rest mass of an electron.
 The rate of this competing process can be expressed just like that for internal conversion.
 $W_{IPF}($E0$)$ is equal to the product of the E0 nuclear transition strength, $\rho^2(\text{E0})$
  multiplied by the electronic factor for internal pair formation, $\Omega_{IPF}(\text{E0})$:
\begin{equation}
 W_{IPF}(\text{E0}) = \rho^2(\text{E0}) \times \Omega_{IPF}(\text{E0}) \, .
\end{equation}

 The total pair conversion transition rate can be expressed as
\begin{equation}
 W_{IPF} = \frac{2\pi}{2J_i + 1} \sum_{i} \sum_{f} |U_{i f}|^2 \delta(W + W' - \omega),
\end{equation}
 where we are averaging over the initial states, and summing over the final states.
 Energy conservation is enforced by the delta function, with the total electron energy ($W^{\prime}$), and the
 total positron energy ($W$), summing to give the transition energy, $\omega$.
 The $U_{i f}$ is the pair transition amplitude which for an E0 transition defined as,
\begin{equation}
 U_{i f} = \sum_{\kappa, \mu} \sum_{\kappa ', \mu '} a^{*}_{\kappa ', \mu '} a_{\kappa, \mu} U^{(L=0)}_{i f}
\end{equation}
 and
\begin{equation}
U_{i f}^{(L=0)} = -e \int_{0}^{\infty} dV_n \rho_n (r_n) \int_{0}^{r_n}
  dr_e [u_{\kappa}(r_e) u'_{\kappa} (r_e) + v_{\kappa}(r_e) v'_{\kappa}(r_e)]
  \Bigg( \frac{1}{r_n} - \frac{1}{r_e} \Bigg) \, .
\end{equation}
 The matrix element $U_{if}^{(L=0)}$ is the same as that for internal conversion as shown in
 Eqn.~(\ref{eqn:E0MatElradial}) except that now the $u_{\kappa}$ and $v_{\kappa}$ are the large and
 small components for the positron, and the $u'_{\kappa}$ and $v'_{\kappa}$ are the large and small
 components for the electron.

 For internal pair formation, we express the electron and positron in the final state as
 Coulomb-distorted plane waves, which are described by spherical Dirac waves in the form of
 Eqn.~(\ref{eqn:diracDef}).
 The decomposition of the Coulomb-distorted planes waves is via the equation
\begin{equation}
 \psi^{(\pm)}_{\textbf{p}, \lambda} = \sum_{\kappa, \mu} a^{(\pm)}_{\kappa \mu}(\Omega, \lambda)
 \chi_{\kappa \mu},
\end{equation}
 where the $a_{\kappa \mu}$ are the expansion coefficients relating a Coulomb-distorted plane wave of definite
 momentum, \textbf{p}, and polarization, $\lambda$, to spherical Dirac waves of definite $\kappa$ and $\mu$.
 The electron and positron are expressed as converging partial waves and described by the $\psi^{(-)}$-type
 wavefunction and the $a^{(-)}$ expansion coefficients.
 The expressions for $a^{(\pm)}_{\kappa \mu}$ and $\psi^{(\pm)}_{\textbf{p}, \lambda}$ can be found in
 \cite{i1990Hofmann}.

 The differential probability of pair emission with respect to the positron energy is given below:
\begin{equation}
  \label{eqn:HofmannE0dPdE}
 \frac{dP_{e^{+}e^{-}}}{dE} = \frac{2\pi}{2J_{i} +1}
 \sum_{M_i, M_f} \sum_{\kappa, \kappa^{\prime}} \sum_{\mu, \mu^{\prime}}
 \sum_{\lambda, \lambda^{\prime}} \int d\Omega \int d\Omega^{\prime} a^{\prime *}_{\kappa^{\prime}
 \mu^{\prime}} a_{\kappa^{\prime} \mu^{\prime}} a^{\prime}_{\kappa \mu} a^*_{\kappa \mu} |U^{(L=0)}_{if}|^2 \, .
\end{equation}

 Following the work of Wilkinson \cite{i1969Wi29}, we can make a number of approximations to simplify
 the problem and make it more tractable.
 We can approximate the electron and positron wavefunctions for a finite-sized nucleus by the pure
 Coulombic wavefunctions for a point nucleus evaluated at the nuclear radius.
 These wavefunctions are be expressed in the notation of beta-decay by the Fermi functions.
 The Fermi functions represent the ratio of the relativistic electron density at the nucleus to the density at
 infinity  \cite{i1966Bahcall}.
 Just as in beta-decay, the nuclear Coulomb effects must be taken into account.
 The nuclear Coulomb field has the effect of inhibiting the emission of a positron, and enhancing
 the emission of an electron.
 Its effect on electron-positron emission is energy dependent and is incorporated into the Fermi functions.
 This also means that we only consider electron and positron final states of $j=\frac{1}{2}$ or $\kappa = \pm 1$.
 This is appropriate as the higher order moments ($r^k$, $k \geq 4$) in the monopole matrix operator are
 negligible \cite{i1956Ch21, i1981So10}.
 The screening of the nuclear charge by the atomic electrons can also be taken into account.
 This is easily done in this formalism by modifying the Fermi functions with the Thomas-Fermi-Dirac model of the
 atom according to the works of Durand, and Bahcall \cite{i1964Durand, i1966Bahcall}.

 By expressing wavefunctions of the electron and positron as pure Coulombic wavefunctions evaluated at
 the nuclear edge,
 we can express the integral as the product of two Fermi functions (one for the positron and  one for the electron).
 The E0 transition rate for pair conversion is evaluated as \cite{i1969Wi29}:
  \begin{equation}
   W_{IPF}(\text{E0}) = \frac{8}{9 \pi }  \frac{e^{4} m_0^{5} c^{4}}{\hbar^{7}}
   | M(\text{E0}) |^{2}    \frac{1}{(1+\gamma)^{2}}
      \int_{1}^{E_{\pi}-1} p_{+} p_{-} (E_{+} E_{-} - \gamma ^{2})
      F(Z,E_{+})F(Z,E_{-}) dE_{+} \, ,
  \label{eqn:W(E0)_pi}
  \end{equation}
  where $e$ is the electron charge, $m_0$ the electron rest mass and $\gamma = (1-\alpha^2 Z^2)^{1/2}$.
   The positron/electron momenta in units of $m_0c$ are denoted $p_{\pm}$ and
  $E_{\pm}$ are the positron/electron total energy in units of $m_0c^2$ with $E_{-}+E_{+}$ equalling
   the transition energy. 
   $F(Z,E_{\pm})$ is the Fermi function and $M(\text{E0})$ is the electric monopole matrix element.

 In order to get an expression for the electronic factor for internal pair formation, we must divide the internal
 pair formation rate, $W($E0$)_{IPF}$ by $\rho^2$ which is related to the monopole matrix element
 by Eqn.~(\ref{eqn:rhoMe0}).
 We then obtain
  \begin{equation}
   \Omega_{IPF}(\text{E0}) = \frac{8}{9 \pi }  \frac{e^{6} R^4 m_0^{5} c^{4}}{\hbar^{7}}  \frac{1}{(1+\gamma)^{2}}
      \int_{1}^{E_{\pi}-1} p_{+} p_{-} (E_{+} E_{-} - \gamma ^{2})
      F(Z,E_{+})F(Z,E_{-}) dE_{+} \, .
  \label{eqn:Omega_pi}
  \end{equation}
 By evaluating this equation we can obtain the electronic
  factor for internal pair formation,
  $\Omega_{IPF}$, as a function of transition energy, and atomic number.

 \subsection{E0 electronic factors and nuclear structure}
 \label{sec:UsingOmgea}

 The electronic factors relate the transition probability for each E0 decay pathway to the nuclear E0
 strength.
 The total E0 transition probability, $W($E0$)$, sums over the different atomic shells and
 processes (electron conversion and internal pair formation) by
 \begin{equation}
   W(\text{E0}) = |\rho(\text{E0})|^2 \times [\Omega_K (\text{E0}) + \Omega_{L1} (\text{E0}) + ... +
   \Omega_{IPF} (\text{E0})] \, ,
   \label{eqn:WE02}
 \end{equation}
 where the transition rate of each component is just
 \begin{equation}
   W_{i}(\text{E0}) = |\rho(\text{E0})|^2 \times \Omega_i (\text{E0}) \, ,
 \end{equation}
and
\begin{equation}
W(\text{E0}) = \sum_{i} W_i(\text{E0}) \, .
\end{equation}

 Experimentally, the monopole strength can be obtained directly if the partial mean life of
 the E0 transition, $\tau($E0$)$, is known. Specifically,
 \begin{equation}
    \rho^{2}(\text{E0}) = \frac{1}{[\Omega_K (\text{E0}) + \Omega_{L1} (\text{E0}) + ... + \Omega_{IPF} (\text{E0})]
     \times \tau(\text{E0})} \, .
 \end{equation}

 For $J^{\pi} \rightarrow J^{\pi}$ transitions in which $J \neq 0$, the transition typically proceeds with a
 mixed M1+E2+E0 character and the $\rho^{2}(E0)$ monopole strength can be extracted as follows.
 The conversion coefficient of this transition involving the i-th subshell or IPF, $\alpha_{i}$, can be written as
 \begin{equation}
   \label{eqn:total_conversion_M1E2E0}
   \alpha_{i} = \frac{I_{i}(\text{M1}) + I_{i}(\text{E2}) + I_{i}(\text{E0})}{I_{\gamma}(\text{M1}) +
   I_{\gamma}(\text{E2})} ,
\end{equation}
 where $I_{i}(\text{ML})$ is the intensity of conversion electrons or pair emission.
 $I_{\gamma}(\text{ML})$ is the gamma-ray emission of multipolarity $\text{ML}$.
 Using the $\delta(\text{E2/M1})$ mixing ratio, the ratio of the E2 to M1 $\gamma$-ray intensities,
 the theoretical $\alpha_{i}(\text{E2})$ and $\alpha_{i}(\text{M1}$) conversion coefficients,
 Eqn.~(\ref{eqn:total_conversion_M1E2E0}) can be written as:
\begin{equation}
\label{eqn:total_conversion_with_delta}
\alpha_{i} = \frac{\alpha_{i}(\text{M1}) + \delta^{2}(\text{E2/M1})[ 1 + q^{2}_{i}(\text{E0/E2})]\alpha_{i}(\text{E2})}{1
 + \delta^{2}(\text{E2/M1})} \, .
\end{equation}
 The $q_{i}^{2}(\text{E0/E2})$ is a quantity introduced by Church, Rose, and Weneser \cite{i1958Church},
  which is the ratio of the E0 and E2 conversion electron (or electron positron pair) intensities:
\begin{equation}
  q^2_i(\text{E0}/\text{E2}) = \frac{I_i(\text{E0})}{I_i(\text{E2})} \, .
\end{equation}%

 If the E2 transition rate, $W_{\gamma}($E2$)$, is known, the $\rho^2($E0$)$ can be obtained
 from the following expression:
\begin{equation}
  \label{eqn:rho2qkWy}
  \rho^2(\text{E0}) = q^2_i (\text{E0}/\text{E2}) \times \frac{\alpha_i (\text{E2})}{\Omega_i (\text{E0})}
  \times W_{\gamma}(\text{E2}) \, ,
\end{equation}
 where as above, the $i$ subscript denotes the atomic shell or pair conversion.

\section{$\Omega($E0$)$ for conversion electrons}
\label{sec:Omega_CE}
\subsection{Numerical calculations}
\label{sec:Calculations_CE}

 The electronic factors, $\Omega_{ce}$, were calculated using a modified version of the \emph{CATAR}
 program \cite{i1975Pa26}.
 The details of the operation of the code can be found in \cite{i1975Pa26}, but a brief outline will
 be given here.
 The program solves the Dirac equation for the bound and free electron states in a central potential
 with a screening function given by the user input.
 These wave functions are used to calculate the electronic factors for internal conversion,
 $\Omega_{ce}$.
 The static nuclear charge distribution of the nucleus is treated as a homogenously charged drop of
  radius $1.20 A^{\frac{1}{3}} \text{fm}$.

 The screening functions used to calculate the tabulated values and accompanying database were
 self-consistent Dirac-Fock-Slater (DFS) functions calculated for each element using a modified version of
 the \emph{HEX} code developed by Liberman, Cromer and Waber \cite{i1971Liberman}.
 The self-consistent potentials were calculated using a modified Kohn-Sham potential outlined in
 \cite{i1971Liberman} and described in \cite{i1970Liberman} and with the application of the so
 called Latter tail correction \cite{i1955Latter}.
 The atomic mass numbers and electronic configurations used in the calculation of each screening
 function were those found in \cite{i2008Ki07}, where applicable, and for elements of $Z \geq 111$,
 the configurations given in \cite{i1971Lu} were used.


 The present calculations of the electronic factors use the so-called ``No-Hole" approximation
 wherein the effect of the atomic vacancy on the atomic potential is disregarded.
 In other words, the self-consistent screening function is left unchanged before and after the conversion event.
 To improve the accuracy, especially for transition energies close to the atomic shell binding energies,
 when the kinetic energy of the outgoing electron is very low, the so-called ``Frozen-Orbitals" approach
 \cite{i2008Ki07} can be used.
 It was determined \cite{i2008Ki07} that for the calculations of the conversion coefficients over the full energy range,
 the difference between the ``No-Hole"  and the ``Frozen-Orbitals" approximations  is about 1.6\%.
 For the E0 transitions of interest, the energy of the continuum electrons is at least of the order
 of 100 keV, so the effect of the hole is expected to be minimal.

 The published version of the \emph{CATAR} program \cite{i1975Pa26} contained some bugs.
 For the present calculations we used the same binding energies as for the \emph{BrIcc} tabulations \cite{i2008Ki07}.
 The interface to provide external screening function has also been modified.
 To calculate the screening function the \emph{HEX} code was also modified by eliminations some bugs along
 with corrections given in \cite{i1971Liberman}.

 For each element of $Z$ from 5 to 126, the self-consistent DFS screening function was calculated using the
 \emph{HEX} code.
 These were then used to calculate the $\Omega_{i}(E0)$ electronic factors for internal conversion electrons
 involving all atomic shells (up to R2) and transition energies from 1 keV above atomic shell binding
 energies to 6000 keV.
 For the N6 shell for $Z=63$, the Q4 shell for $Z=122$, and the O8 shell for $Z=125$ and $126$,
 the calculations could not be performed due to instability in the numerical solution of the bound
 electronic wave function.

\begin{figure}[t]
 \begin{center}
    	\includegraphics[width=0.8\linewidth]{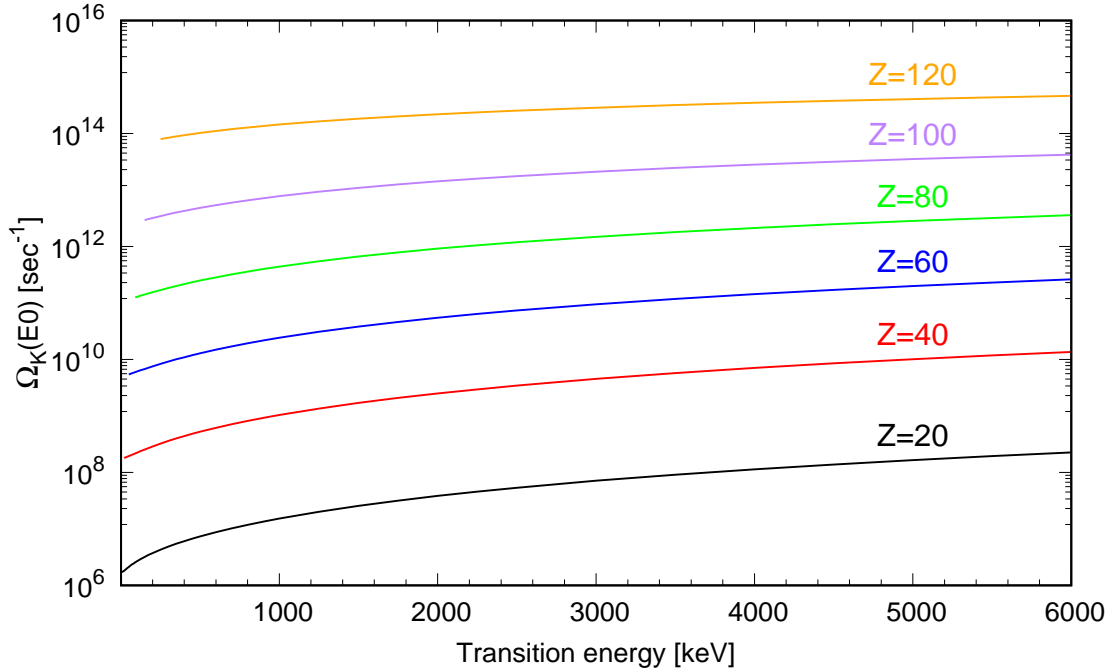}
 \end{center}
    \vspace*{-10pt}
	\caption{Plot of $\Omega_{K}($E0$)$ electronic factors for the K shell as a function of energy and for atomic numbers
     $Z$ = 20, 40, 60, 80, 100 and 120.}
	\label{fig:OmegaValues}
\end{figure}

\begin{figure}
    \vspace*{-50pt}
    \begin{center}
	\includegraphics[width=0.7\textwidth]{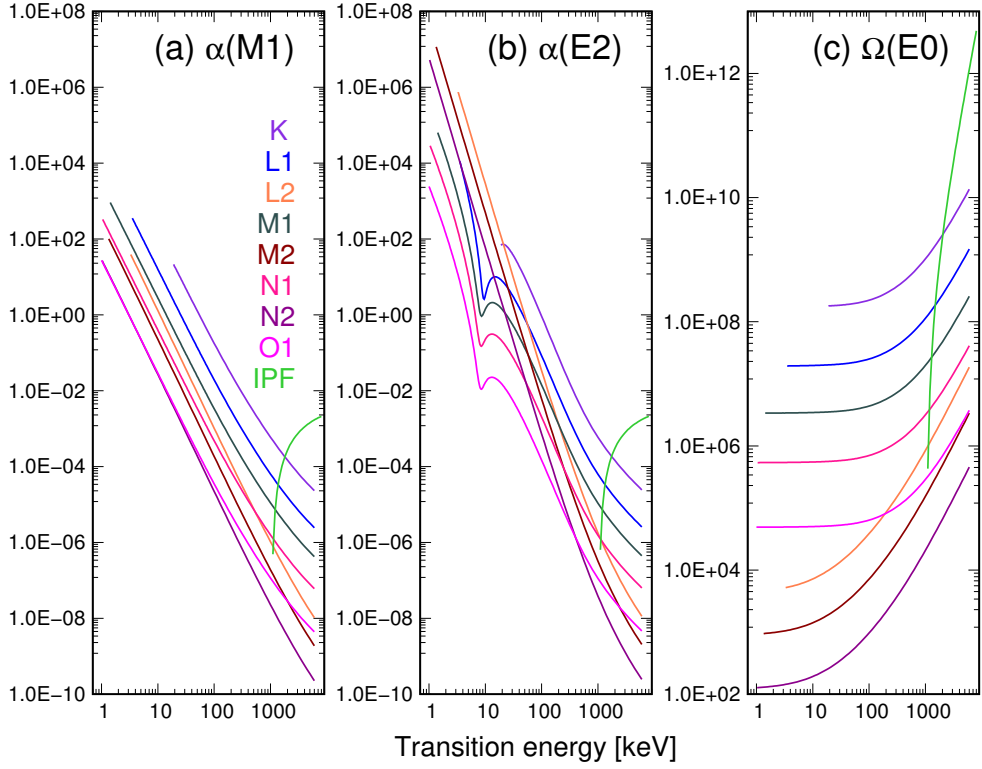}
    \end{center}
    \vspace*{-10pt}
	\caption{Plot of theoretical internal conversion coefficient for
    \emph{(a)} M1, \emph{(b)} E2
    multipolarity and
    \emph{(c)} E0 electronic factors for Z=40 as a function of transition energy.
    Note the logarithmic scale on both axes.
    For clarity, only the K, L1, L2, M1, M2 and O1 shells, as well as internal pair formation are shown.
    $\alpha($M1$)$ and $\alpha($E2$)$ conversion coefficients are from \emph{BrIcc} \cite{i2008Ki07}.
    The $\Omega($E0$)$ values are from the present tabulation. }
	\label{fig:Z40_M1E2E0}
\end{figure}

 Figure~\ref{fig:OmegaValues} shows the $\Omega_{K}(\text{E0})$ values as a function of transition energy
 for selected atomic numbers.
 The electronic factors are increasing as a function of atomic number $Z$ and transition energy.
 The same general dependence was found for all other shells.
 Compared to the K-shell conversion coefficients, $\alpha_K$ (see Fig.~\ref{fig:Z40_M1E2E0}), the energy
 dependence of the $\Omega_{CE}($E0$)$ values is opposite to that of the conversion coefficients.
This energy dependence is valid for all atomic shells. On the other hand, the $Z$-dependence is the same for
 both the E0 electronic factors and the internal conversion coefficients.

\begin{figure}[t]
 \begin{center}
 	\includegraphics[width=0.8\linewidth]{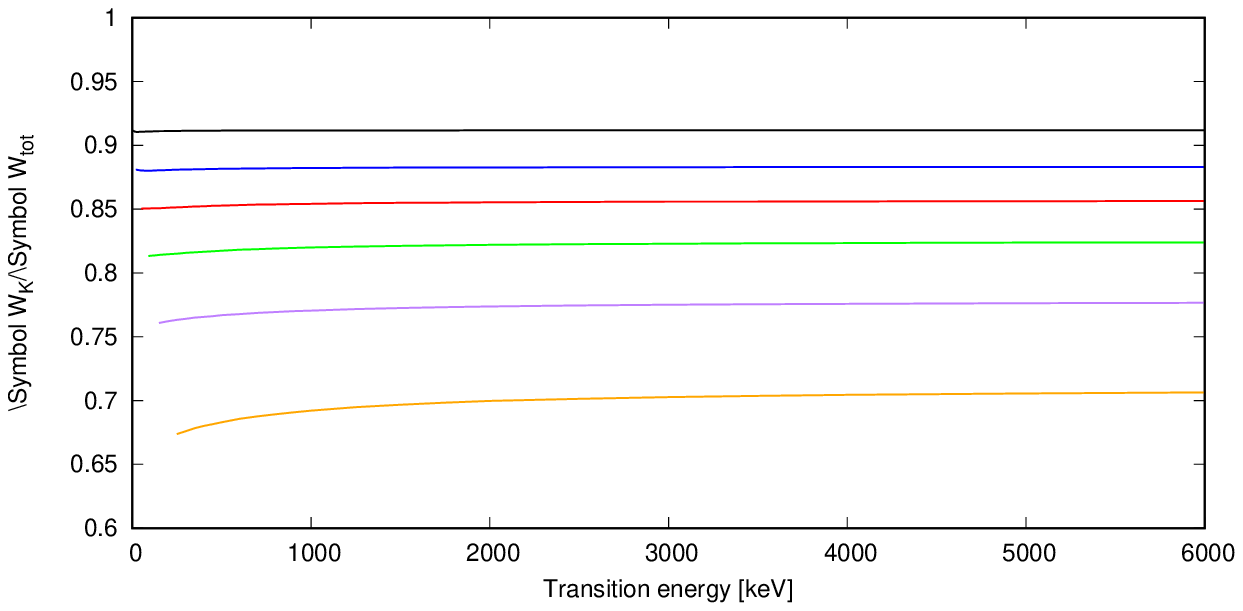}
 \end{center}
    \vspace*{-10pt}
	\caption{Plot of the ${\Omega_K}/{\Omega_{tot,CE}}$ ratio as a function of transition energy and for
    atomic number Z=20, 40, 60, 80, 100 and 120.
    Note, that ${\Omega_{tot,CE}}$ only contains contributions from electron conversion.}
	\label{fig:Omega_KTot}
\end{figure}

 Figure~\ref{fig:Omega_KTot} shows the ratio of the $\Omega_{K}($E0$)$ to the $\Omega_{tot,CE}($E0$)$ electronic factors.
 Here $\Omega_{tot,CE}($E0$)$ only contains contributions from electron conversion.
 It is notable that the ratio has very little energy dependence, however it decreases with atomic number $Z$.

\subsection{Comparison with previous $\Omega_{CE}($E0$)$  tabulations}
\label{sec:OmegaCe_comparison}

 The previous E0 electronic factor tabulations are far from complete as the different tabulations used different
 physical assumptions, and covered different regions in terms of transition energy, atomic number, and atomic shell.
 Representative values obtained from the previous tabulations are compared in Table~\ref{tab:OmegaComp_CE}
 with values from this new tabulation.
 In each case $\Omega_{i}($E0$)$ values were obtained by numerical interpolations.

 The 1969 tabulation by Hager and Seltzer \cite{i1969Ha61} (\textit{HsOmg}) was carried out for every fourth $Z$
 between $Z=30$ and $Z=102$.
 The values for the K, L1, and L2 shells were calculated, with transition energies ranging from 6 keV above
 the threshold to 1500 keV for the K shell, and 1100 keV for the L1 and L2 shells.
 The finite size of the nucleus was approximated as a Fermi distribution,
 and the wave functions were solved via a relativistic Hartree-Fock-Slater approach with the
 use of the Kohn-Sham corrections to the Slater approach \cite{i1965Kohn}.

 The second tabulation, by Bell \textit{et al.}  \cite{i1970Be87} (\textit{BeOmg}) in 1970,
 included every second $Z$ from $Z=40$ to 102 and for the K, L1, and L2 shells.
 The electronic factors were calculated by modifying the analytic point-nucleus wave functions
 from Church and Weneser \cite{i1956Ch21} to account for the finite nuclear size and atomic screening
 effects and covered an electron energy range from 51 keV to 2.5 MeV.

 The third tabulation was by Passoja and Salonen in 1986 \cite{i1986PaZM} (\textit{PaOmg}).
 This tabulation covered every even $Z$ from $Z=8$ to $Z=40$.
 The electronic factors for electron conversion were only from the K shell and
 calculated for transition energies from 511 keV to 12.8 MeV.
 The effects of the finite-sized nucleus and atomic screening were not taken into account and
 this leads to an overestimation of the $\Omega_{K}($E0$)$ electronic factor.
 This is visible in Table~\ref{tab:OmegaComp_CE} at $Z=40$ compared to the other tabulations.

 The agreement between \textit{HsOmg}, \textit{BeOmg} and the present \textit{CATAR} calculations
 is reasonably good, except for L2 shell and $Z=40$, where the \textit{BeOmg} values are systematically
 lower.
 The present calculations are the only one for atomic shells higher than L2.
 Our calculation also treat the effects of the finite nuclear size  as well as atomic screening
 consistently for all elements.

\newpage

\renewcommand{\arraystretch}{0.7}
\begin{longtable}[t]{cccccccc}
\caption{\label{tab:OmegaComp_CE}Comparison of $\Omega_{CE}($E0$)$ electronic
    factor tabulations for conversion electrons.} \vspace*{-18pt} \\
  \multicolumn{1}{c}{Transition energy} &
  \multicolumn{1}{c}{$Z$  } &
  \multicolumn{1}{c}{Shell  } &
  \multicolumn{4}{c}{$\Omega_{CE}($E0$)$ [1/sec]  } \\ \cline{4-7}
  \multicolumn{1}{c}{[keV]} &
  \multicolumn{1}{c}{  } &
  \multicolumn{1}{c}{  } &
  \multicolumn{1}{c}{\emph{HsOmg}}&
  \multicolumn{1}{c}{\emph{BeOmg}}&
  \multicolumn{1}{c}{\emph{PaOmg}}&
  \multicolumn{1}{c}{\emph{CATAR}}\\
  \multicolumn{1}{c}{  } &
  \multicolumn{1}{c}{  } &
  \multicolumn{1}{c}{  } &
  \multicolumn{1}{c}{ \cite{i1969Ha61} }&
  \multicolumn{1}{c}{ \cite{i1970Be87} }&
  \multicolumn{1}{c}{ \cite{i1986PaZM} }&
  \multicolumn{1}{c}{ This work }\\ \hline
  \endfirsthead
\caption{Continued} \vspace*{-18pt} \\
  \multicolumn{1}{c}{Transition energy} &
  \multicolumn{1}{c}{$Z$  } &
  \multicolumn{1}{c}{Shell  } &
  \multicolumn{4}{c}{$\Omega_{CE}($E0$)$ [1/sec]  } \\ \cline{4-7}
  \multicolumn{1}{c}{[keV]} &
  \multicolumn{1}{c}{  } &
  \multicolumn{1}{c}{  } &
  \multicolumn{1}{c}{\emph{HsOmg}}&
  \multicolumn{1}{c}{\emph{BeOmg}}&
  \multicolumn{1}{c}{\emph{PaOmg}}&
  \multicolumn{1}{c}{\emph{CATAR}}\\
  \multicolumn{1}{c}{  } &
  \multicolumn{1}{c}{  } &
  \multicolumn{1}{c}{  } &
  \multicolumn{1}{c}{ \cite{i1969Ha61} }&
  \multicolumn{1}{c}{ \cite{i1970Be87} }&
  \multicolumn{1}{c}{ \cite{i1986PaZM} }&
  \multicolumn{1}{c}{ This work }\\ \hline
  \endhead

   600 &  10 &  K  &          &          & 3.05E+5  & 2.77E+5  \\
       &     &  L1 &          &          &          & 1.58E+4  \\
       &     &  L2 &          &          &          & 2.79E+0  \\
  1000 &     &  K  &          &          & 5.37E+5  & 4.90E+5  \\
       &     & L1  &          &          &          & 2.80E+4  \\
       &     & L2  &          &          &          & 6.61E+0  \\
  1500 &     &  K  &          &          & 9.10E+5  & 8.30E+5  \\
       &     & L1  &          &          &          & 4.74E+4  \\
       &     & L2  &          &          &          & 1.34E+1  \\
       &     &     &          &          &          &          \\
   600 &  40 &  K  & 6.20E+8  & 6.23E+8  & 6.62E+8  & 6.15E+8  \\
       &     & L1  & 6.88E+7  & 6.61E+7  &          & 6.80E+7  \\
       &     & L2  & 3.89E+5  & 3.28E+5  &          & 3.84E+5  \\
  1000 &     &  K  & 1.05E+9  & 1.06E+9  & 1.09E+9  & 1.04E+9  \\
       &     & L1  & 1.15E+8  & 1.11E+8  &          & 1.14E+8  \\
       &     & L2  & 8.64E+5  & 7.28E+5  &          & 8.49E+5  \\
  1500 &     &  K  & 1.71E+9  & 1.74E+9  & 1.72E+9  & 1.70E+9  \\
       &     & L1  &          & 1.83E+8  &          & 1.86E+8  \\
       &     & L2  &          & 1.43E+6  &          & 1.65E+6  \\
       &     &     &          &          &          &          \\
   600 &  82 &  K  & 3.78E+11 & 3.77E+11 &          & 3.81E+11 \\
       &     & L1  & 6.37E+10 & 6.28E+10 &          & 6.41E+10 \\
       &     & L2  & 2.35E+9  & 2.30E+9  &          & 2.36E+9  \\
  1000 &     &  K  & 5.81E+11 & 5.87E+11 &          & 5.86E+11 \\
       &     & L1  & 9.55E+10 & 9.54E+10 &          & 9.65E+10 \\
       &     & L2  & 4.43E+9  & 4.38E+9  &          & 4.46E+9  \\
  1500 &     &  K  & 8.70E+11 & 8.96E+11 &          & 8.79E+11 \\
       &     & L1  & 1.41E+11 & 1.43E+11 &          & 1.43E+11 \\
       &     & L2  & 7.56E+9  & 7.57E+9  &          & 7.64E+9  \\
\hline
\end{longtable}

\section{$\Omega($E0$)$ for pair conversion}
\label{sec:Omega_IPF}

\subsection{Numerical calculations}
\label{sec:Calculations_IPF}

 Our main aim is to present complete tables of $\Omega_{IPF}($E0$)$ to be used with \emph{BrIcc}.
 The electronic factors, $\Omega_{IPF}($E0$)$, were calculated using \emph{WspOmega}, a new computational
 tool developed from the work of Wilkinson \cite{i1969Wi29}.
 $\Omega_{IPF}($E0$)$ values have been calculated for even $Z$ from 4 to 100
 using the atomic masses of the most abundant isotopes adopted for \emph{BrIcc} \cite{i2008Ki07}.
 The range of nuclear transition energies was between 1100 keV and 8000 keV similar to
 the $L >0$ pair conversion coefficients tabulations \cite{i1979Sc31,i1996Ho21}.
 The smooth dependence of $\Omega_{IPF}($E0$)$ on $Z$ and transition energy makes it easy to obtain values
  for odd $Z$ by interpolating from neighbouring even $Z$ values.

 The electronic factors for internal pair formation were calculated by numerical integration of
  Eqn.~\ref{eqn:Omega_pi} with the Simpson's rule.
 The calculation of the Fermi function is based on Eqn.~7 from \cite{i1999Firestone}.
 The effect of the atomic screening is corrected for using a Thomas-Fermi-Dirac statistical model of
 the atom following the works of Durand and Bahcall \cite{i1964Durand, i1966Bahcall}.

 Only electron and positron states with $j=\frac{1}{2}$ or $\kappa = \pm 1$ were considered.
 This is because the effects of the higher order moments ($r^k$, $k \geq 4$) in the monopole matrix operator are
 negligible \cite{i1956Ch21, i1981So10}.
 The nuclear Coulomb effects have been taken into account but not the effect of the finite nuclear size.
 The Fermi functions correspond to Dirac wavefunctions for a point source nucleus, evaluated at the
 nuclear edge \cite{i1969Wi29}.
 The effects of the finite nuclear size are expected to be important at high-$Z$ systems and at transition
  energies close to the threshold of 1.022 MeV.
 Wilkinson \cite{i1969Wi29} has estimated that these effects could alter the $\Omega_{IPF}($E0$)$ by not more than 10\%.

\subsection{Comparison with previous $\Omega_{IPF}($E0$)$ calculations}

 There are only a handful of theoretical calculations of numerical $\Omega_{IPF}($E0$)$ values in
 the literature:
 the work of Lombard, Perdrisat and Brunner \cite{i1968Lo16}, Passoja and Salonen \cite{i1986PaZM},
 and Soff \cite{i1981So10}.
 There is only one published data table -- the work of Passoja and Salonen -- which only covers
  nuclei of $Z<40$.
  The other two papers only contain $\Omega_{IPF}($E0$)$ values for a small selection of nuclear transitions.

Table~\ref{tab:OmegaComp_IPF} compares the ratios of $\Omega($E0$)$ values for the K-shell to
  pair conversion obtained from  previous publications and the present calculations to available
  experimental intensity ratios of K-shell to pair formation.

 Lombard, Perdrisat and Brunner \cite{i1968Lo16} carried out so-called ``exact" calculations of the
 internal pair formation electronic factor using a modified Born approximation and relativistic electron wave functions for a point nuclear
 charge calculated by Bhalla and Rose \cite{i1961BhallaRose}.
 The published tables were for transition energies of 2.5, 3.0, 4.0, and 5.0 $m_ec^{2}$ or for
 1278, 1533, 2044, 2555 keV, and only for $Z$ = 0, 34, 48, 58, and 82.
 The finite nuclear size, and atomic screening was not taken into account.
 Their calculated values only agree with experiment at very low atomic numbers.
 the difference to the experiments and to the other calculations is quite significant for $Z=20$ and above.

 Passoja and Salonen \cite{i1986PaZM} calculated the internal pair formation  $\Omega_{IPF}($E0$)$
 electronic factors for atomic numbers $Z$ = 8--40 and transition energies of 1430 to 12775 keV.
 To our knowledge, this is the only data table on $\Omega_{IPF}($E0$)$; it was published in 1986 as an
 internal report of the University of Jyv\"{a}skyl\"{a}.
 This calculation is based on Born approximation with corrections for the nuclear Coulomb effects from
  Wilkinson \cite{i1969Wi29} along with  wavefunctions for a point nucleus.
  Atomic screening and finite nuclear size were both neglected in the calculations.

 A more realistic model of the partial pair emission probability, $d\Omega_{IPF}($E0$)/dE$  was developed
 in 1981 by Soff, Schl\"uter and Greiner \cite{i1981So10}.
 Their model utilize relativistic electron and positron wavefunctions \cite{i1979Soff}, which are
 solutions of the Dirac equation with a point nucleus potential.
 To include the effect of the finite nuclear size, the density of the electron states are included
 approximately, however the screening of atomic electrons is neglected.
 These calculations only provide numerical values of $\Omega_{IPF}(E0)/\Omega_{K}(E0)$ ratios and
 according to our knowledge the code is not available.


 Table~\ref{tab:OmegaComp_IPF} compares the ratios of $\Omega($E0$)$ values for the K-shell to the
 pair conversion calculated with the various models.
 The available experimental data for E0 intensity ratios between the K-shell and internal pair formation is also shown.
 The agreement between Passoja \cite{i1986PaZM}, Soff \cite{i1981So10} and the present calculations is
 reasonably good.
 Note that in Table~\ref{tab:OmegaComp_IPF}, the K/IPF value for the 6048.2 keV transitions in $^{16}$O
 calculated in this work is significantly lower than in the works of Lombard \textit{et al.},
 Passoja and Salonen, and Soff.
 This difference may arise because the $\Omega_K$ values of Lombard \textit{et al.}, Passoja and Salonen,
 and Soff were calculated with point nuclear wavefunctions and no atomic screening.
 As seen in Table~\ref{tab:OmegaComp_CE}, the $\Omega_K$ values of Passoja and Salonen are characteristically
 larger than the values from the present calculations using \emph{CATAR}, the Hager and Seltzer tabulation,
 and the Bell \textit{et al.} tabulation.
 This will inflate the values of the K/IPF $\Omega ($E0$)$ ratios.

\setlength{\tabcolsep}{5pt}
\renewcommand{\arraystretch}{0.8}
\begin{longtable}{cllccllll}
\caption{Experimental data and theoretical values of K-shell electron to internal pair conversion\\
    probability ratios for E0 transitions.}
\label{tab:OmegaComp_IPF} \vspace*{-20pt} \\
  \multicolumn{1}{c}{Nucleus} &
  \multicolumn{1}{c}{Transition} &
  \multicolumn{7}{c}{$\Omega_{K}($E0$) / \Omega_{IPF}($E0$)$}\\ \cline{3-9}
  \multicolumn{1}{c}{ } &
  \multicolumn{1}{c}{Energy   } &
  \multicolumn{3}{c}{Experiment} &
  \multicolumn{1}{c}{Lombard } &
  \multicolumn{1}{c}{Passoja } &
  \multicolumn{1}{c}{Soff } &
  \multicolumn{1}{c}{This work} \\
  \multicolumn{1}{c}{ } &
  \multicolumn{1}{c}{ [keV]}&
  \multicolumn{3}{c}{} &
  \multicolumn{1}{c}{\cite{i1968Lo16} } &
  \multicolumn{1}{c}{\cite{i1986PaZM} } &
  \multicolumn{1}{c}{\cite{i1981So10} } &
  \multicolumn{1}{c}{ }\\ \hline
$^{16}$O                                         &  
6048.2                                           &  
4.00E-5 &    \emph{(46)} &   \cite{i1963Le06}    &  
3.92E-5                                          &  
3.82E-5                                          &  
3.8E-5                                           &  
3.45E-5                                          \\ 
$^{40}$Ca                                        &  
3352.6                                           &  
6.94E-3    & \emph{(20)}  &   \cite{i1962Ne02}   &  
6.0E-3                                           &  
7.16E-3                                          &  
7.16E-3                                          &  
6.86E-3                                          \\ 
$^{42}$Ca                                        &  
1837.3                                           &  
0.111      & \emph{(22)}  &   \cite{i1961Be19}   &  
0.072                                            &  
0.139                                            &  
0.139                                            &  
0.133                                            \\ 
$^{54}$Fe                                        &  
2561.3                                           &  
0.053      & \emph{(14)}  &   \cite{i1986Pa19}   &  
                                                 &  
0.0598                                           &  
                                                 &  
0.0575                                           \\ 
 &
 &
0.053      & \emph{(3)}  &   \cite{i2018Eriksen}   &  
                                                 &  
0.0598                                           &  
                                                 &  
0.0575                                           \\ 
$^{60}$Ni                                        &  
2284.87                                          &  
0.130      & \emph{(28)} &   \cite{i1981Pa10}    &  
                                                 &  
0.135                                            &  
                                                 &  
0.135                                            \\ 
$^{64}$Zn                                        &  
1910                                             &  
0.46       & \emph{(7)}  &   \cite{i1985Pa07}    &  
                                                 &  
0.472                                            &  
                                                 &  
0.475                                            \\ 
$^{70}$Ge                                        &  
2307.1                                           &  
0.20       & \emph{(8)}  &   \cite{i1985Pa15}    &  
                                                 &  
0.212                                            &  
                                                 &  
0.207                                            \\ 
$^{90}$Zr                                        &  
1760.70                                          &  
3.0       & \emph{(11)}  &   \cite{i1957Yu06}     &  
1.42                                             &  
2.48                                             &  
2.52                                             &  
2.48                                             \\ 
                                                 &  
                                                 &  
2.38       & \emph{(8)}  &   \cite{i1962Ne02}    &  
1.42                                             &  
2.48                                             &  
2.52                                             &  
2.48                                             \\ 
                                                 &  
                                                 &  
2.28       & \emph{(32)} &   \cite{i1980Pa_thesis}    &  
1.42                                             &  
2.48                                             &  
2.52                                             &  
2.48                                             \\ 
$^{140}$Ce                                       &  
1903.5                                           &  
6.3      &              &   \cite{i1960An05}     &  
5.22                                             &  
                                                 &  
6.9                                              &  
7.23                                            \\ 
$^{214}$Po                                       &  
1416                                           &  
440-625      &              &   \cite{i1948Gei,i1940Alikhanov}     &  
148                                             &  
                                                 &  
388.6                                          &  
414                                            \\ 

\hline
\end{longtable}


\section{Numerical data tables}

 The new $\Omega_{CE}($E0$)$ and $\Omega_{IPF}($E0$)$ tabulations have been combined with
 our previous calculations of internal conversion electron coefficients \cite{i2008Ki07,i2012Ki04},
 as well as pair conversion coefficients calculated by Schulter \cite{i1979Sc31} and Hofmann \cite{i1996Ho21}
 as part of the \emph{BrIcc} database (version 3.1).
 Details are given in Table~\ref{tab:IccTabulations}.
 For additional details about the \emph{BrIcc} database, see \cite{i2008Ki07}.
 The new data tables were assembled as a single file, containing probabilities of E0, E1 to E5 and M1 to M5
 multipole orders, for electron and pair-conversion.

\renewcommand{\arraystretch}{0.8}
\begin{longtable}{llclll}
\caption{Tabulations of internal conversion coefficients and
    electronic factors available in \emph{BrIcc} version 3.1.}
    \label{tab:IccTabulations} \vspace*{-20pt} \\
\multicolumn{1}{c}{Data Table } &
\multicolumn{1}{c}{ Reference} &
\multicolumn{1}{c}{ $Z$ } &
\multicolumn{1}{c}{ Shell$^{a}$ } &
\multicolumn{1}{c}{ L }&
\multicolumn{1}{c}{ $E_{\gamma}^{b}$ [keV]}\\
\hline
\multicolumn{6}{c}{\textbf{Conversion electrons}}\\
\multicolumn{6}{c}{$\alpha_{IC}$ Conversion coefficient }\\
\hline
\multicolumn{1}{l}{\textbf{BrIccFO}  } &
\multicolumn{1}{l}{\cite{i2002Ba85,i2008Ki07}} &
\multicolumn{1}{c}{5 -- 110} &
\multicolumn{1}{c}{K - Q1} &
\multicolumn{1}{c}{1 -- 5}&
\multicolumn{1}{c}{$\epsilon_{ic}$+1 -- 6000} \\
\multicolumn{1}{l}{  } &
\multicolumn{1}{l}{\cite{i2002Ba85,i2012Ki04}} &
\multicolumn{1}{c}{111 -- 126} &
\multicolumn{1}{c}{K - R2} &
\multicolumn{1}{c}{1 -- 5}&
\multicolumn{1}{c}{$\epsilon_{ic}$+1 -- 6000} \\
\hline
\multicolumn{6}{l}{    } \\
\multicolumn{6}{c}{$\Omega_{IC}(E0)$ electronic factors }\\
\hline
\multicolumn{1}{l}{\textbf{CATAR}  } &
\multicolumn{1}{l}{Present work} &
\multicolumn{1}{c}{5 -- 126} &
\multicolumn{1}{c}{K - R2} &
\multicolumn{1}{c}{0}&
\multicolumn{1}{c}{$\epsilon_{ic}$+1 -- 6000} \\
\hline
\multicolumn{6}{l}{    } \\
\multicolumn{6}{c}{\textbf{Electron-positron pair emission} }\\
\multicolumn{6}{c}{$\alpha_{\pi}$ Conversion coefficient }\\
\hline
\multicolumn{1}{l}{\textbf{ScPcc} }&
\multicolumn{1}{l}{ \cite{i1979Sc31}} &
\multicolumn{1}{c}{0 -- 49}&
\multicolumn{1}{c}{IPF }&
\multicolumn{1}{c}{1 -- 3}&
\multicolumn{1}{c}{1100 -- 8000}\\
\multicolumn{1}{l}{\textbf{HoPcc} }&
\multicolumn{1}{l}{ \cite{i1996Ho21}} &
\multicolumn{1}{c}{50 -- 100}&
\multicolumn{1}{c}{IPF }&
\multicolumn{1}{c}{1 -- 3}&
\multicolumn{1}{c}{1100 -- 8000}\\
\hline
\multicolumn{6}{l}{    } \\
\multicolumn{6}{c}{$\Omega_{IPF}(E0)$ electronic factor}\\
\hline
\multicolumn{1}{l}{\textbf{WspOmega} }&
\multicolumn{1}{l}{ Present work} &
\multicolumn{1}{c}{4 -- 100 (even)}&
\multicolumn{1}{c}{IPF }&
\multicolumn{1}{c}{0}&
\multicolumn{1}{c}{1100 -- 8000}\\
\hline
\multicolumn{6}{l}{$^{a}$ \hspace*{5mm} IPF labels internal pair formation in the atomic shell field. See also Table~\ref{tab:AtomicShells}.}\\
\multicolumn{6}{l}{$^{b}$ \hspace*{5mm} $\epsilon_{ic}$ is the binding energy for the ith atomic shell.}\\
\end{longtable}

 To obtain values for a given $Z$, transition energy, atomic shell, or multipolarity, the use of a cubic spline interpolation is recommended.
 The data tables listed in Table~\ref{tab:IccTabulations} are distributed in a fixed record size
 binary file, designed for modern computers and operating systems.
 Each record is 52 bytes long, can hold up to 12 single-precision numbers: transition energy,
 $\Omega($E0$)$  electronic factor and ten $\alpha$ conversion coefficients for E1-E5 and M1-M5 multipolarities.
 The last 4 bytes are reserved for future developments.
 For each element, the data tables are generated for up to 40 atomic shells.
 Electron-positron pair conversion is considered for storage as one of the atomic shells as listed in Table~\ref{tab:AtomicShells}.
 Within a given shell, conversion coefficients were calculated for a number of transition energies.
 The numbers and values of the energies were chosen to minimize the uncertainty from the
 interpolation procedure to be well below 0.3\%  \cite{i2008Ki07}.

 The first record of the data file has the version of the data base, followed by 126 blocks of 41 records for each $Z$.
 The conversion data starts at the 5169th record.
 Full details of the data file, including the list of the records, are given in the \textit{ReadMe} file available with the auxiliary files of this paper.


\renewcommand{\arraystretch}{0.8}
\begin{longtable}{lll}
\caption{Atomic shells used in the data tables. }
    \label{tab:AtomicShells} \vspace*{-28pt} \\
\multicolumn{1}{c}{ Index } &
\multicolumn{1}{c}{ Shell} &
\multicolumn{1}{c}{ Atomic orbits}  \\
\hline
 1      & K    & $1s_{1/2} $ \\
 2-4    & L1-3 & $2s_{1/2}, \, 2p_{1/2}, \, 2p_{3/2} $ \\
 5-9    & M1-5 & $3s_{1/2}, \, 3p_{1/2}, \, 3p_{3/2}, \, 3d_{3/2}, \, 3d_{5/2} $ \\
 10-16  & N1-7 & $4s_{1/2}, \, 4p_{1/2}, \, 4p_{3/2}, \, 4d_{3/2}, \, 4d_{5/2}, \, 4f_{5/2}, \, 4f_{7/2} $ \\
 17-25  & O1-9 & $5s_{1/2}, \, 5p_{1/2}, \, 5p_{3/2}, \, 5d_{3/2}, \, 5d_{5/2}, \, 5f_{5/2}, \, 5f_{7/2} , \, 5g_{7/2}, \, 5g_{9/2}$ \\
 26-32  & P1-7 & $6s_{1/2}, \, 6p_{1/2}, \, 6p_{3/2}, \, 6d_{3/2}, \, 6d_{5/2}, \, 6f_{5/2}, \, 6f_{7/2} $ \\
 33-37  & Q1-5 & $7s_{1/2}, \, 7p_{1/2}, \, 7p_{3/2}, \, 7d_{3/2}, \, 7d_{5/2} $ \\
 38-39  & R1-2 & $8s_{1/2}, \, 8p_{1/2} $ \\
 40     & $\pi$ & Internal electron-positron pair formation, IPF \\
 \hline
 \end{longtable}



\section{Comparison with experiments}
\label{sec:ExpTheor}

 The basic relationship between E0 transitions rates and the electronic factor, $\Omega(\text{E0})$, is
 given in Eqn.~(\ref{eqn:omegarho}).
 While the absolute E0 transition rates can be obtained from lifetime measurements or high-energy
 electron-scattering experiments, neither $\rho($E0$)$ nor $\Omega($E0$)$ can be determined
 independently from experiments.
 Accurate calculated values of $\Omega($E0$)$ are therefore essential to deduce $\rho($E0$)$
 from absolute E0 transition rates.
 While $\Omega($E0$)$ cannot be experimentally determined, the ratio of E0 internal conversion
 electron and/or E0 pair conversion intensities is equal to the ratio of the corresponding
 $\Omega($E0$)$ electronic factors.

 In Table~\ref{tab:E0IccRatio}, 128 experimental $\Omega($E0$)$ ratios are given, covering a wide
 range of atomic numbers (from $Z=8$ to $Z=98$), transition energies (from 171.077 keV in
 $^{114}$Cd to 6048.2 keV in $^{16}$O), and involving K-- to N--shells and internal pair formation.

 The corresponding theoretical $\Omega_{CE}($E0$)$ and $\Omega_{IPF}($E0$)$ values have been
 evaluated from the present calculations and listed in the ``Theory" column of Table~\ref{tab:E0IccRatio}.
 The difference between the experimental intensity ratios ($\Omega^{\text{ratio}}_{\text{Exp}}$) and theoretical
  electronic factor ratios ($\Omega^{\text{ratio}}_{\text{Theory}}$),
 $\Delta \Omega^{\text{ratio}}$,  is in percent and defined as:
 \begin{equation}
 \label{eqn:ExpTheorDif}
   \Delta \Omega^{\text{ratio}}  = \frac{100 \times [\Omega^{\text{ratio}}_{\text{Exp}} -
   \Omega^{\text{ratio}}_{\text{Theory}}]}
                                           {\Omega^{\text{ratio}}_{\text{Theory}}} \, .
 \end{equation}
 The uncertainty given for $\Omega^{\text{ratio}}$ only  contains the contributions from the experimental values.
 References to the experimental data are given with their NSR \cite{iNSR} key numbers.

 There is a significant variation in the experimental values and their uncertainties.
 In a few cases only a limit is known.
 Most of the conversion electron data is on K/L ratios and there are only a few measurements on sub-shell ratios.
To quantify the difference between the experimental and theoretical ratios, data points from Table~\ref{tab:E0IccRatio} with relative uncertainty $\leq25\%$ were selected.

\begin{figure}[t]
\begin{center}
\vspace*{-20pt}
\includegraphics[angle=0,width=13.3cm]{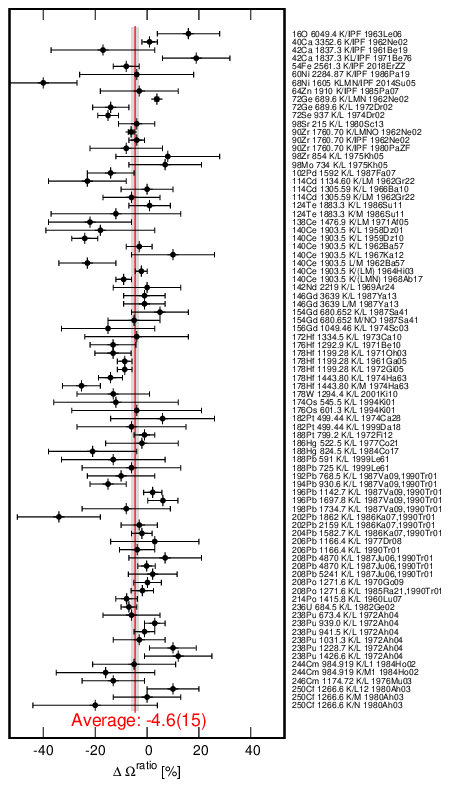}
\vspace*{-30pt}
\end{center}
\caption{Percentage differences between experimental E0 intensity ratios and theoretical $\Omega($E0$)$
 electronic factor ratios, $\Delta \Omega^{\text{ratio}}$.
 See Eqn.~(\ref{eqn:ExpTheorDif}) for definition.
 Experimental data with less than $\pm25\%$ relative uncertainty taken from Table~\ref{tab:E0IccRatio}.
  The average difference of $\Delta \Omega^{\text{ratio}} \, =\, -4.6(15)$\% is indicated by a shaded area. }
\label{fig:ExpTheor}
\end{figure}

 These 83 data points are plotted in Fig.~\ref{fig:ExpTheor}, where the horizontal axis is the percentage difference between
 theory and experiment, as defined in Eqn.~(\ref{eqn:ExpTheorDif}).
 The data points are plotted by decreasing mass numbers.
 References are given on the right-hand side.
Eight of the data points are K/IPF ratios which correspond to the following E0 transitions: 6048.2 keV $^{16}$O, 3352.6 keV $^{40}$Ca, 1837.3 keV $^{42}$Ca, 2561.3 keV $^{54}$Fe, 2284.87 keV $^{60}$Ni,
1910 keV $^{64}$Zn, and 1760.70 keV $^{90}$Zr.

 Considering electron conversion only, our calculations show (see Fig.~\ref{fig:Omega_KTot}) that  between $Z=32$ (Ge)
 and $Z= 98$ (Cf) the contribution of K conversion changes from $\sim 90\%$ to $\sim 80\%$.
 A visual examination of Fig.~\ref{fig:ExpTheor} indicates that the difference between experiment and theory
 does not have any significant $Z$ dependence. The eight K/IPF ratios included also appear independent of Z.

 To determine the average difference, we have used AveTools \cite{iAveTools}, which combines five different
 statistical methods to calculate averages of experimental data with uncertainties:
 (a) Limitation of Relative Statistical Weight (LWM) \cite{i1992Ra08},
 (b) Normalised Residual Method (NRM) \cite{i1992Ra08},
 (c) Rajeval Technique (RT) \cite{i1992Ra08},
 (d) Bootstrap method (BS) \cite{i2002He06},
 (e) Mandel-Paule approach (MP) \cite{i2011Ch22}.
 A robust answer would be reached if all these methods gave the same answer, or the averages with their uncertainties
 are overlapping.
 The most precise experimental value is the K/L+M+N+O ratio of 8.70 \emph{(6)} (0.69\% relative uncertainty)
 of the 689.6 keV E0 transition in $^{72}$Ge reported by Nessin et al. \cite{i1962Ne02}.
 It is more than 5$\sigma$ away from the corresponding theoretical value of  8.38.
 This point was marked as outlier by the LWM, NRM, RT methods, so its uncertainty was increased to 2\%.
 The K to L ratio of the 1199.28 keV E0 transition in $^{178}$Hf \cite{i1974Ha63} has been reported
 with about 9\% precision, however the deviation from the theoretical ratio is nearly 50\%.
 This point has been excluded from the present analysis.

 \begin{longtable}{lcc}
\caption{Average $\Delta \Omega^{\text{ratio}}$ differences for 83 E0 transitions shown in
  Fig.~\ref{fig:ExpTheor}. \\The averages have been determined using five statistical approaches.}
 \label{tab:AverExpTheor}
 \vspace*{-8mm} \\
\multicolumn{1}{c}{ Method}&
\multicolumn{1}{c}{ $\Delta \Omega^{\text{ratio}}$ (\%)}&
\multicolumn{1}{c}{ Reduced $\chi^2$}\\ \hline
 Limitation of Relative Statistical Weight \cite{i1992Ra08}    & -4.6\emph{(15)} & 1.56 \\
 Normalised Residual Method \cite{i1992Ra08}                   & -4.9\emph{(7)} & 1.40 \\
 Rajeval Technique \cite{i1992Ra08}                            & -4.9\emph{(6)} & 1.30 \\
 Bootstrap method \cite{i2002He06}                             & -4.7\emph{(12)} & 1.51 \\
 Mandel-Paule Approach \cite{i2011Ch22}                        & -5\emph{(4)} & 1.51 \\
 \hline
 \end{longtable}

 Average values for these $\Delta \Omega^{\text{ratio}}$ differences  are given
 in Table~\ref{tab:AverExpTheor}.
 All five methods indicate that the various ratios of E0 transition intensities are overestimated in the
 model by about 5\%.
 The same data set was also compared with $\Omega($E0$)$ values from the old \emph{BrIcc} database
 using $\Omega_{CE}($E0$)$ values from Hager and Seltzer \cite{i1969Ha61},
  Bell \textit{et al.} \cite{i1970Be87}, and Passoja \cite{i1986PaZM}
  along with $\Omega_{IPF}($E0$)$ values from Passoja \cite{i1986PaZM}.
 The average difference was slightly greater at 5.5(17)\%.

 The average differences for the 10 experimental ratios involving IPF are:
 LWM: -2.6\emph{(18)}\%,
 NRM: -2.5\emph{(19)}\%,
 RT: -2.1\emph{(19)}\%,
 BS: -4\emph{(4)}\%,
 MP: -3.8\emph{(10)}\%.
 These values are overlapping and consistent with each other.
 There appears to be consistent overestimation of the ratio of K-shell conversion to internal pair
  formation of around 2.5\%.
  Compared to the average difference of all data in Table~\ref{tab:AverExpTheor}, the theory
  appears to more accurate for the K-shell to pair formation ratio.
  More experimental data with increased accuracy is needed to establish a firm
  conclusion on the uncertainty of the K-shell and IPF $\Omega($E0$)$ values.

 The average difference between experimental and theoretical E0 ratios in Table~\ref{tab:E0IccRatio} is
 dominated by K/L ratios and more experimental data are needed to examine the model predictions for
 outer shells.
 However, it is possible that the -4.6\emph{(15)}\% difference could be due to a systematic
 underestimation of the electronic factors for the L and higher shells compared to the K-shell.
 The theoretical K/IPF ratios are significantly more consistent with the experimental values,
 suggesting that the K and IPF $\Omega$ values are more accurate than the values for the higher shells.
 The formation region for internal conversion of E0 transitions is wholly within the nuclear volume
 and is purely a penetration effect.
 This is in contrast to internal conversion  processes of $L>0$ where the formation region is
 outside the nuclear volume \cite{i1970Band_NP}.
 Thus the $\Omega($E0$)$ values are very sensitive to the electron wavefunctions inside the
 nucleus and so depend very strongly on the particulars of the calculated wavefunctions
 and electronic configurations.
 The outer shells are even more sensitive due to the effect of atomic screening on the
 wavefunctions from the inner shells.
 From the present work, shown in Table~\ref{tab:AverExpTheor}, we have adopted a 5\%
 uncertainty on the calculated $\Omega_{CE}($E0$)$ and $\Omega_{IPF}($E0$)$ values.

\section{Using $\Omega_{CE}$ and $\Omega_{IPF}$ electronic factors for analyzing E0 transitions}

\noindent
 Eqn.~(\ref{eqn:WE02}) describes the relation between E0 transition rate, the monopole strength parameter,
 $\rho($E0$)$, and the electronic factor, $\Omega($E0$)$.
 The conversion electron intensity for a particular shell (\emph{i}) or for pair conversion ($\pi$) is proportional
 to $\Omega($E0$)$:
 \begin{equation}
   I_{i,IPF} \propto \Omega_{i,IPF}(\text{E0}) \, .
 \label{eqn:IntE0}
 \end{equation}
 Using the present table one can evaluate the complete set of $\Omega_{i,IPF}($E0$)$ values.
 Here we provide some examples on the use of $\Omega($E0$)$ electronic factors to analyse
 pure E0 and mixed E0+E2+M1 transitions.

 \subsection{$^{42}$Ca 1837.3 keV E0}

 The ratio of the K conversion electron and pair conversion of the pure E0 transition in $^{42}$Ca
 was measured by Benzer-Koller \emph{et al.} \cite{i1961Be19} (Table~\ref{tableI}): $I_K / I_{IPF} = 0.111 (22)$.
 This experimental ratio 17(20)\% is lower than the calculated one using the present tabulation.
 Belyaev, Vasilenko and Kaminker \cite{i1971Be76} measured the  pair production to K+L
 conversion electron ratio, $I_{IPF}/I_{K+L}=6.1 \, (8)$, which is 13(15)\% higher than
 the calculated one.
 The 1837.31 keV $0^+$ state also decays with a 312.60 keV E2 transition.
 The ratio of the K-shell conversion electron intensities of the E2 and E0 transitions is reported by
 Benzer-Koller \emph{et al.} \cite{i1961Be19}, $I_K(E2)/I_K(E0)=1.03(10)$.
 To evaluate the ratio of the total E0 to E2 intensities here we only make use of the experimental
  $I_K(E2)/I_K(E0)$ value and the theoretical conversion coefficients and E0 electronic factors:
$\alpha_{K}(E2)=0.00310(5)$,
$\alpha_{tot}(E2)=0.00340(5)$,
$\Omega_{IPF}($E0$)=2.54(13)E+8$
and
$\Omega_{tot}($E0$)=2.91(15)E+8$.
 From these data the E0/E2 branching ratio was deduced as:
 \begin{equation}
  \frac{I_{tot}(E2)}{I_{tot}(\text{E0})}  = \frac{I_K(\text{E2})}{I_K(\text{E0})} \times \frac{1+\alpha_{tot}(\text{E2})}{\alpha_K(\text{E2})}
  \times \frac{I_K(\text{E0})}{I_{IPF}(\text{E0})} \times \frac{\Omega_{IPF}(\text{E0})}{\Omega_{tot}(\text{E0})}
    =  32(7) \label{eqn:TIE2_TIE0}
\end{equation}
 Using Eqn.~\ref{eqn:TIE2_TIE0} and the adopted half-life of the 1837.31 keV $0^+$ state \cite{i2016Ch23},
 $T_{1/2} = 387(6)$ ps the $\rho^2($E0$)$ can be obtained as:
 \begin{equation}
  \rho^2(\text{E0}) =  \frac{ln (2)}{\Omega_{tot}(\text{E0}) \times (1+\frac{I_{tot}(\text{E2})}{I_{tot}(\text{E0})})\times T_{1/2}}
    =  0.19(4) \, . \label{eqn:rho2E0}
\end{equation}
 In comparison, the 2005 evaluation of E0 transition strengths by Kib\'edi and Spear \cite{i2005Ki02}
 gave $\rho^2($E0$)=0.140(12)$.
 This value is based on the $I_{tot}(E0)/I_{\gamma}=2.05 (17)/100.0(18)$ from the 2001 ENSDF evaluation
 \cite{i2001Si10}.

\subsection{$^{238}$U 921.10 keV E0+E2+M1}

\begin{figure}[tb]
    \begin{center}
	\includegraphics[width=0.45\textwidth,angle=0]{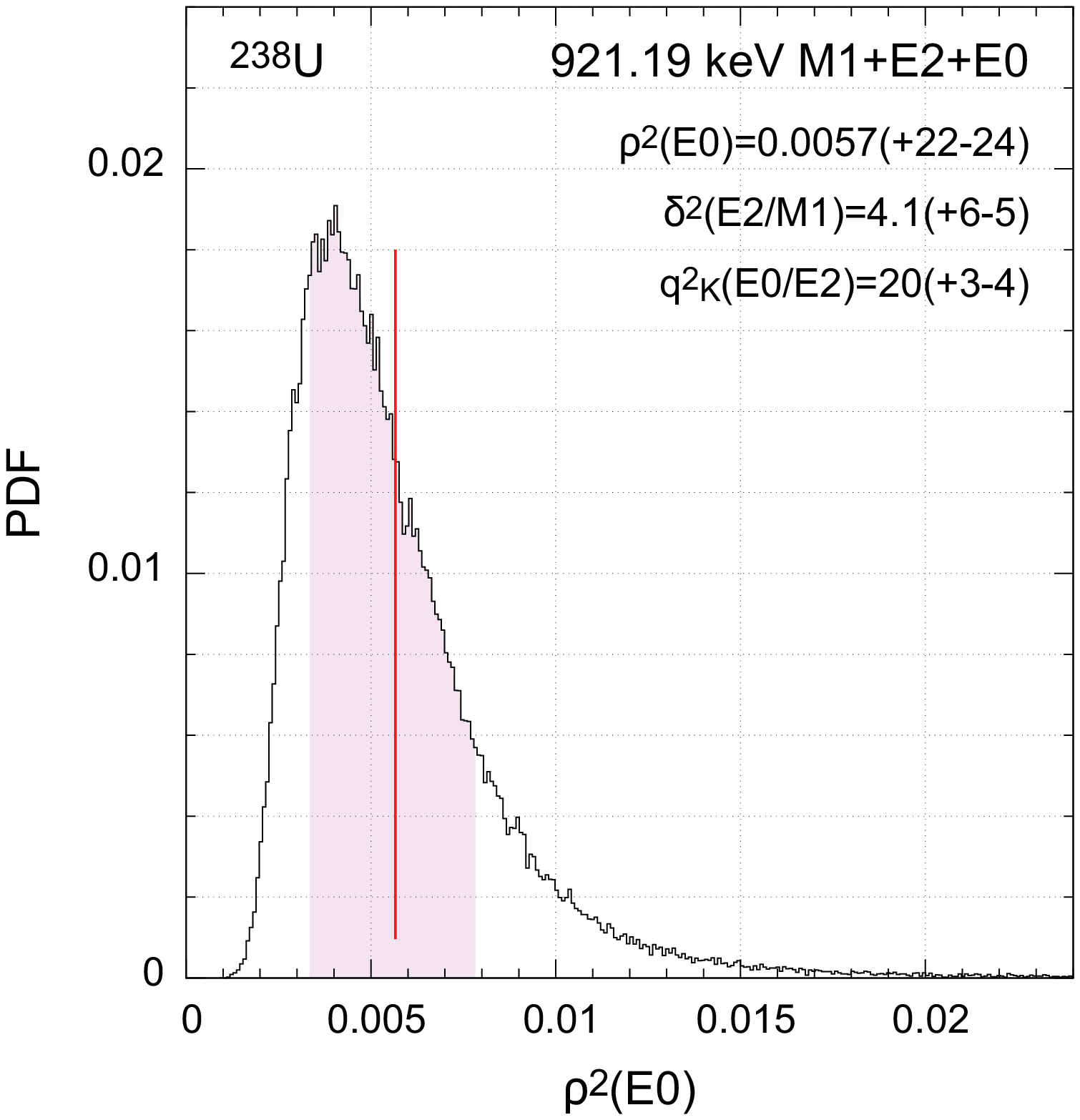}
    \end{center}
    \vspace*{-20pt}
	\caption{Plot of probability density function of $\rho^2($E0$)$ evaluated using
    Eqn.~(\ref{eqn:238U_rho}).
    The red line indicates the mean value of $\rho^2($E0$)$ and the shaded area corresponds to
    the upper and lower 1 $\sigma$ boundaries.}
	\label{fig:238U_921M1E2E0_RHO2E0}
\end{figure}


\begin{longtable}{lcccl}
\caption{Transitions from the 966.13 keV $2^+$ state in $^{238}$U$^{a}$.
\label{tab:238U_DecayData}}
\vspace{-30pt} \\
\multicolumn{1}{c}{$E_{\gamma}$ [keV]} &
\multicolumn{1}{c}{ $I_{\gamma}$} &
\multicolumn{1}{c}{ Multipolarity}&
\multicolumn{1}{c}{ $J^{\pi}_{f}$}&
\multicolumn{1}{c}{Comments} \\ \hline
\endhead
234.5(10)   &   13.9(14) & E1 & $3^{-}$         &                           \\
286.3(10)   &   8.1(7)   & E1    & $1^{-}$     &                           \\
818.06(13)  &   100(4)   & E2  & $4^{+}$       &                           \\
921.19(3)   &    60(3)   & E0+E2+M1 & $2^{+}$  & $\delta^2($E2/M1$)$=+4.1$^{+0.6}_{-0.5}$\\
            &            &           &  & $\alpha_K(exp)$=0.191(30)   \\
966.9(3)    &   27.3(14) & E2 & $0^{+}$         &                           \\
 \hline
\multicolumn{5}{l}{$^{a}$ Data from \cite{i2015Br06}}\\
 \end{longtable}

 The decay data of the 966.13 keV $2^+$ state in $^{238}$U was taken from the recent ENSDF evaluation of
 Browne and Tuli \cite{i2015Br06} and is listed in Table~\ref{tab:238U_DecayData}.
 The 921.19 keV transition is a $2^+ \rightarrow 2^+$ transition.
  The experimental  K-conversion coefficient, $\alpha_K(exp)$=0.191(30) is larger than the M1
  ($\alpha_K($M1$)$=0.0343) or  E2 ($\alpha_K($E2$)$=0.00887) values, indicating an E0 admixture.
 Using Eqn.~(\ref{eqn:total_conversion_with_delta}) the E0/E2 mixing ratio can be extracted as:
 \begin{eqnarray}
   q^2_K(\text{E0/E2}) & = & \frac{\alpha_K(exp) \times (1+\delta^2(\text{E2/M1}))-\alpha_K(\text{M1})}
   {\delta^2(\text{E2/M1}) \times\alpha_K(\text{E2})} - 1 \\
    & = & 20^{+3}_{-4} \, . \nonumber
 \end{eqnarray}
 The resulting E0/E2 mixing ratio is slightly asymmetric, however the
 total conversion coefficient, $\alpha_{tot}=0.26(4)$ deduced from the theoretical $\alpha_{i}($M1$)$
  and $\alpha_{i}($E2$)$ and $\Omega_{i}($E0$)$ values, using $\delta($E2$/$M1$)$ and
  $q_{i}($E0$/$E2$)$ mixing ratios (\emph{i} = K, L1, L2, etc. atomic shells) has a  symmetric uncertainty.
 It should be noted that the value of the E0/E2 mixing ratio is not the same for the various atomic shells
 or for pair conversion.
 Using the $\Omega($E0$)$ and $\alpha($E2$)$ values $q^{2}_{j}($E0$/$E2$)$ can be obtained from $q^{2}_{i}($E0$/$E2$)$
 as:
 \begin{equation}
   q^{2}_{j}(\text{E0}/\text{E2}) = q^{2}_{i}(\text{E0}/\text{E2}) \times \frac{\Omega_{j}(\text{E0})}{\Omega_{i}(\text{E0})} \times
   \frac{\alpha_{i}(\text{E2})}{\alpha_{j}(\text{E2})} \, .
 \end{equation}

 Using Eqn.~\ref{eqn:TIE2_TIE0} the E0 strength parameter:
 \begin{eqnarray}
   \rho^2(\text{E0}) & = & q^2_K(\text{E0}/\text{E2}) \times \alpha_K(\text{E2}) \times
   \frac{I_{\gamma}(\text{E2})} {\sum I_{tot}} \times  \frac{ln(2)}{T_{1/2}}\\
   & = &  0.0057^{+22}_{-24} \, , \nonumber
   \label{eqn:238U_rho}
 \end{eqnarray}
 where $I_{\gamma}($E2$)$ and $\alpha_K($E2$)$ are the gamma-ray intensity and K-shell conversion coefficient
 of the 921.19 keV transition.
 Other parameters to deduce the total EM radiation intensity de-exciting the 966.13 keV level, $\sum I_{tot}$,
 were taken from  Table~\ref{tab:238U_DecayData}.
 The level half-life of $T_{1/2}= 2.4^{+17}_{-7}$ ps is very asymmetric, therefore a
 Monte-Carlo uncertainty propagation program \cite{i2019Combes} was used to evaluate the mean value
 and asymmetric uncertainties from the probability density function (PDF) of $\rho^{2}($E0$)$ shown in
 Fig.~\ref{fig:238U_921M1E2E0_RHO2E0}.
 The present value of $\rho^2($E0$)=0.0056^{+22}_{-24}$ is lower than the $\rho^2($E0$)=0.0099(18)$
 reported by G\'acsi \emph{et al.} \cite{i2001Ga55}.
 A large part of the difference can be attributed to the fact that in \cite{i2001Ga55} the contribution of the
 M1 multipolarity to the K-shell conversion coefficient has not been taken into account.

\section{Conclusion}
  Here we report on the calculation of E0 electronic factors for internal conversion and
  electron-positron pair conversion.
  These tables will replace the current ones in \emph{BrIcc} \cite{i2008Ki07}.

  For the electron conversion a modified version of the \emph{CATAR} program \cite{i1975Pa26} was
  used. $\Omega_{CE}($E0$)$ values are presented for all $nS$ and $nP$ atomic shells in the $Z = 5$ to $126$ elements
  and  covering the same energy  range as for the conversion coefficients in \emph{BrIcc}.
  For pair conversion a new code has been developed using Wilkinson's formulation \cite{i1969Wi29} for elements of even
  $Z = 4$ to $100$ and 1100 to 8000 keV transition energies.

  In Table~\ref{tab:OmegaComp_IPF} the calculated $\Omega_{CE}($E0$)$ and $\Omega_{IPF}($E0$)$ values are
  compared with available theoretical calculations and a satisfactory agreement was found.
  To our knowledge, there is no experimental data is available to make a direct comparison of
   $\Omega_{CE}($E0$)$ or $\Omega_{IPF}($E0$)$ values.
  In Table~\ref{tab:E0IccRatio}, 128 ratios of $\Omega($E0$)$ values are compared with theoretical values.
  Fig.~\ref{fig:ExpTheor} shows the percentage differences between experiment and theory.
  No obvious \emph{Z} or mass dependence is visible in the figure.
  The average difference between the experimental and theoretical values is $-4.6(15)\%$, indicating
  that the theoretical ratios are overestimating experiments.
  Further theoretical work is needed to improve the calculations of $\Omega(E0)$.

  Based on the above analysis, a general $5\%$ uncertainty was adopted for the accuracy of the
  present numerical tables.
  A new data table has been built using electron and pair conversion coefficients for $L>0$ multipolarities from
  our previous calculations \cite{i2008Ki07,i2012Ki04} and combined with the present calculations
  for E0 transitions.
  The binary table is available as an auxiliary file to this publication and will be
   available in the new version of \emph{BrIcc}.

\label{sec:Conlusion}

\ack
 We would like to thank G.~Gosselin, V.~Meot and M.~Pascal (CEA, DAM, DIF, Arpajon F-91297, France),
 who havefig:ExpTheor carried out the initial calculations of electronic factors for conversion electrons.
 This work was supported by the Australian Research Council Discovery Grants DP140102986
 and DP170101673.




\section*{Table 1.\label{tab:E0IccRatio_Explanation}
  Ratio of experimental conversion electron (CE) and electron-positron pair conversion
  intensities for E0 transition  in even--even atomic nuclei.}
\begin{tabular*}{0.95\textwidth}{@{}@{\extracolsep{\fill}}lp{5.5in}@{}}
\multicolumn{2}{p{0.95\textwidth}}{(Throughout this table,
	italicized numbers refer to the uncertainties in the last
	digits of the quoted values.)}\\

Nuclide	& The even $Z$, even $N$ nuclide studied\\

$E_{\text{tran}}$
	& Energy of the E0 transition in keV either from the Adopted levels and gammas
	of the Nuclear Data Sheet or from cited literature\\

Shell
	& Atomic electron shell (K,L,M, etc), or electron-positron pair (IPF)\\

$\Omega^{\text{ratio}}_{\text{Exp}}$	
    & Ratio of the experimental pure E0 intensities\\

$\Omega^{\text{ratio}}_{\text{Theory}}$	
    & Ratio of the calculated  $\Omega($E0$)$ electronic factors from the present tabulation\\

Difference
 & Relative percentage difference between the experimental E0 intensity ratio,
  $\Omega^{\text{ratio}}_{\text{Exp}}$ and the calculated 
  $\Omega($E0$)$ electronic factor ratio, $\Omega^{\text{ratio}}_{\text{Theory}}$. 
  See Eqn.~\ref{eqn:ExpTheorDif}\\

References
	& NSR \cite{iNSR} key numbers of the source of the experimental intensity ratios.

\end{tabular*}
\label{tableI}

\datatables            
\renewcommand{\arraystretch}{1.5}


\setlength{\LTleft}{0pt}
\setlength{\LTright}{0pt}


\setlength{\tabcolsep}{0.5\tabcolsep}

\footnotesize 



\newpage
\begin{theDTbibliography}{1956He83}
 \bibitem[\href{http://www.nndc.bnl.gov/nsr/nsrlink.jsp?1957Yuasa,B}{1957Yuasa}]{1957Yu06}
  T. Yuasa, \emph{et al.}
  \emph{``D\'esint\'egration de $^{90}$Y. Transition $0^{+} \rightarrow 0^{+}$  $^{90}$Zr"},
   J.~Phys.~Radium \textbf{18} (1957) 498.
 \bibitem[\href{http://www.nndc.bnl.gov/nsr/nsrlink.jsp?1958Dz01,B}{1958Dz01}]{1958Dz01}
   B.S.~D\v{z}elepow, Yu~V.~Kholnov and V.P.~Prikhodtseva,
   \emph{``A $0^{+} \rightarrow 0^{+}$ transition in $^{140}$Ce"},
    Nucl. Phys. \textbf{9} (1958) 665.
 \bibitem[\href{http://www.nndc.bnl.gov/nsr/nsrlink.jsp?1959Dz10,B}{1959Dz10}]{1959Dz10}
   B.S. Dzhelepov, I.F. Uchevatkin and S.A. Shestopalova,
   \emph{``$0^{+} - 0^{+}$ Transition in the $^{140}$Pr $\rightarrow$ $^{140}$Ce Decay Scheme"},
    Soviet Phys. JETP \textbf{10} (1960)  611.
 \bibitem[\href{http://www.nndc.bnl.gov/nsr/nsrlink.jsp?1960An05,B}{1960An05}]{1960An05}
    S.F. Antonova, \emph{et al.},
    \emph{``Investigation of the gamma spectrum of $^{140}$Ce"},
     Sov. J. of Physics JETP \textbf{11} (1960) 554.
 \bibitem[\href{http://www.nndc.bnl.gov/nsr/nsrlink.jsp?1960Fr06,B}{1960Fr06}]{1960Fr06}
   J. Fr\'ana, I. R(ezanka, M. Vobeck\'y and V. Hu\v{s}\'ak,
   \emph{``Neutron-deficient isotopes of terbium with half-life of 18 hours"},
   Czech. J. Phys. \textbf{10} (1960) 692.
 \bibitem[\href{http://www.nndc.bnl.gov/nsr/nsrlink.jsp?1960Lu07,B}{1960Lu07}]{1960Lu07}
   G.~Luhrs and C. Mayer-B\"oricke,
   \emph{``Conversion Line Spectrum of $^{214}Po$ (RaC')"},
    Z.~Naturforsch \textbf{15a} (1960) 939.
 \bibitem[\href{http://www.nndc.bnl.gov/nsr/nsrlink.jsp?1961Be19,B}{1961Be19}]{1961Be19}
    N. Benczer Koller, M. Nessin and T.H. Kruse,
   \emph{ Nuclear Electric Monopole Transition in $^{42}$Ca},
    Phys. Rev.  \textbf{123} (1961) 262.
 \bibitem[\href{http://www.nndc.bnl.gov/nsr/nsrlink.jsp?1961Ga05,B}{1961Ga05}]{1961Ga05}
    C.J. Gallagher, \emph{et al.},
   \emph{``Intrinsic Excited States in $^{178}$Hf Populated by the Allowed Decay of 9.3-min $^{178}$Ta"},
    Phys. Rev. \textbf{122} (1961) 1590.
 \bibitem[\href{http://www.nndc.bnl.gov/nsr/nsrlink.jsp?1961Ha23,B}{1961Ha23}]{1961Ha23}
    B. Harmatz, T.H. Handley and J.W. Mihelich,
    \emph{``Nuclear Levels in a Number of Even-Even Rare Earths (150 $<$ A $<$ 184)"},
    Phys. Rev. \textbf{123} (1961) 1758.
 \bibitem[\href{http://www.nndc.bnl.gov/nsr/nsrlink.jsp?1962Ba57,B}{1962Ba57}]{1962Ba57}
    V.A. Balalaev, \emph{et al.},
    \emph{``Refinement of the Information on the $O^{+} \rightarrow O^{+}$ Transition in $^{140}$Ce"},
    Soviet Phys. JETP \textbf{16} (1963) 1425.
\bibitem[\href{http://www.nndc.bnl.gov/nsr/nsrlink.jsp?1962Bi16,B}{1962Bi16}]{1962Bi16}
    E. Bieber,  Till v. Egidy and Otto W.B. Schult,
   \emph{``Das Neutroneneinfangspektrum von $^{150}$Sm"},
    Z. f\"ur Phys. \textbf{A170} (1962) 465.
\bibitem[\href{http://www.nndc.bnl.gov/nsr/nsrlink.jsp?1962Gr22,B}{1962Gr22}]{1962Gr22}
    L.V.~Groshev, \emph{et al.},
    \emph{``Gamma--Ray and Internal Conversion Electron Spectra of the $^{113}$Cd(n,$\gamma$)$^{114}$Cd Reaction"},
     Columbia Tech. Transl. \textbf{26}  (1963) 987.
\bibitem[\href{http://www.nndc.bnl.gov/nsr/nsrlink.jsp?1962Ne02,B}{1962Ne02}]{1962Ne02}
    M. Nessin, T.H. Kruse and K.E. Eklund,
    \emph{``Nuclear electric Monopole Transitions in $^{16}$O, $^{40}$Ca, $^{72}$Ge, and $^{90}$Zr"},
    Phys. Rev. \textbf{125} (1962) 639.
 \bibitem[\href{http://www.nndc.bnl.gov/nsr/nsrlink.jsp?1963Le06,B}{1963Le06}]{1963Le06}
    Y.K.~Lee, L.W. Mo and C.S. Wu,
    \emph{``Observation of the internal conversion line from the 6.052--MeV level in $^{16}$O"},
     Phys. Rev. Lett. \textbf{10} (1963) 258.
 \bibitem[\href{http://www.nndc.bnl.gov/nsr/nsrlink.jsp?1963Le17,B}{1963Le17}]{1963Le17}
    C.M. Lederer,
    \emph{``The Structure of heavy Nuclei: a study of very weak alpha branching"},
    Univ.California (1963), UCRL-11028.
\bibitem[\href{http://www.nndc.bnl.gov/nsr/nsrlink.jsp?1964Hi03,B}{1964Hi03}]{1964Hi03}
    K. Hisatake, \emph{et al.},
    \emph{``The $0^{+}$ excited state of $^{140}$Ce studied by the decay of $^{140}$Pr"},
    Nucl. Phys. \textbf{56}  (1964) 625.
\bibitem[\href{http://www.nndc.bnl.gov/nsr/nsrlink.jsp?1964Pe17,B}{1964Pe17}]{1964Pe17}
    N.F. Peek, J.A. Jungerman and C.G. Patten,
    \emph{``Nuclear energy levels of $^{156}$Gd as populated by beta emission from $^{156}$Eu"},
    Phys. Rev. \textbf{136} (1964) B330.
\bibitem[\href{http://www.nndc.bnl.gov/nsr/nsrlink.jsp?1966Ba10,B}{1966Ba10}]{1966Ba10}
    A. B\"acklin, N.E. Holmberg and G. B\"ackstr\"om,
    \emph{``Internal conversion study of $^{113}$Cd(n,$\gamma$)$^{114}$Cd"},
    Nucl. Phys. \textbf{80} (1966) 154.
 \bibitem[\href{http://www.nndc.bnl.gov/nsr/nsrlink.jsp?1966Gr04,B}{1966Gr04}]{1966Gr04}
    R. Graetzer, \emph{et al.},
   \emph{``Vibrational states of deformed nuclei populated by the $(p,2n)$ reaction"},
    Nucl. Phys. \textbf{76} (1966) 1.
\bibitem[\href{http://www.nndc.bnl.gov/nsr/nsrlink.jsp?1967Ka12,B}{1967Ka12}]{1967Ka12}
    S.E. Karlsson, \emph{et al.},
    \emph{``The decay of $^{140}$La"},
    Nucl. Phys. \textbf{A100}  (1967) 113.
\bibitem[\href{http://www.nndc.bnl.gov/nsr/nsrlink.jsp?1967Ko15,B}{1967Ko15}]{1967Ko15}
    J. Kormicki, \emph{et al.},
   \emph{``Newly observed gamma transitions in $^{151}$Tb and $^{152}$Tb decay"},
    Acta Phys. Pol. \textbf{31} (1967) 317.
\bibitem[\href{http://www.nndc.bnl.gov/nsr/nsrlink.jsp?1967Ma29,B}{1967Ma29}]{1967Ma29}
    G.~Malmsten, O.~Nilsson and I.~Andersson,
    \emph{``On the decay of $^{152}$Eu"},
    Arkiv Fysik \textbf{33} (1967) 361.
\bibitem[\href{http://www.nndc.bnl.gov/nsr/nsrlink.jsp?1967Vr04,B}{1967Vr04}]{1967Vr04}
    J. Vrzal, \emph{et al.},
    \emph{``System of collective I$^{\pi}$ = 0$^{+}$, K = 0 Levels in $^{164}$Er"},
     Bull. Acad. Sci. USSR, Phys. Ser. \textbf{31} (1968) 599.
\bibitem[\href{http://www.nndc.bnl.gov/nsr/nsrlink.jsp?1968Ab17,B}{1968Ab17}]{1968Ab17}
    L.N. Abesalashvili, \emph{et al.},
    \emph{``Excitation of levels of the 'semimagic' nucleus $^{140}$Ce in decay of $^{143}$Pr"},
     Bull. Acad. Sci. USSR, Phys. Ser. \textbf{32} (1969) 730.
\bibitem[\href{http://www.nndc.bnl.gov/nsr/nsrlink.jsp?1969Ar24,B}{1969Ar24}]{1969Ar24}
    R. Arlt, \emph{et al.},
   \emph{``Levels in $^{142}$Nd"},
    Bull. Acad. Sci. USSR, Phys. Ser. \textbf{33}   (1970) 1505.
\bibitem[\href{http://www.nndc.bnl.gov/nsr/nsrlink.jsp?1970Ag05,B}{1970Ag05}]{1970Ag05}
    P.G.E.~Reid, M.~\v{S}ott, N.J.~Stone,
    \emph{``A study of the magnetic dipole moments of $^{192}$Ir and $^{194}$Ir and the decay scheme of $^{192}$Ir by nuclear orientation"},
    Nucl. Phys. \textbf{A129} (1969) 273.
\bibitem[\href{http://www.nndc.bnl.gov/nsr/nsrlink.jsp?1970Go09,B}{1970Go09}]{1970Go09}
    L.H. Goldman, \emph{et al.},
    \emph{``E0 Transitions from $0_{2}^{+} \rightarrow 0_{1}^{+}$ states in the Z=82 region"},
    Phys. Rev. \textbf{C1} (1970) 1781.
\bibitem[\href{http://www.nndc.bnl.gov/nsr/nsrlink.jsp?1971Af05,B}{1971Af05}]{1971Af05}
    V.P. Afanasev, \emph{et al.},
    \emph{``Decay of $^{138}$Nd, $^{138g}$Pr and $^{138}$*Pr"},
    Izv. Akad. Nauk SSSR: Ser. Fiz. \textbf{35} (1971) 1603.
\bibitem[\href{http://www.nndc.bnl.gov/nsr/nsrlink.jsp?1971Be10,B}{1971Be10}]{1971Be10}
    F.M. Bernthal, J.O. Rasmussen and J.M. Hollander,
    \emph{``Decays of $^{176}$Ta, $^{176}$Lu, and $^{176m}$Lu to levels in $^{176}$Hf"},
    Phys. Rev. \textbf{C3} (1971) 1294.
\bibitem[\href{http://www.nndc.bnl.gov/nsr/nsrlink.jsp?1971Oh03,B}{1971Oh03}]{1971Oh03}
    H. Ohlsson and  J. Gizon,
    \emph{``Transitions E0 et E0+M1+E2 de $^{178}$Hf"},
     C.R. Acad. Sci., Ser. B \textbf{272} (1971) 633.
\bibitem[\href{http://www.nndc.bnl.gov/nsr/nsrlink.jsp?1972Ah04,B}{1972Ah04}]{1972Ah04}
    I. Ahmad, \emph{et al.},
    \emph{``Electron Capture Decay of $^{238}$Am and electric monopole transitions in
     $^{238}$Pu"},
     Nucl. Phys. \textbf{A186} (1972) 620.
\bibitem[\href{http://www.nndc.bnl.gov/nsr/nsrlink.jsp?1972Fi12,B}{1972Fi12}]{1972Fi12}
    M.~Finger, \emph{et al.},
    \emph{``Properties of low--lying levels in the even platinum nuclei $(182 < A < 192)$"},
    Nucl. Phys., \textbf{A188} (1972) 369.
\bibitem[\href{http://www.nndc.bnl.gov/nsr/nsrlink.jsp?1972Gi05,B}{1972Gi05}]{1972Gi05}
    J. Gizon, \emph{et al.},
    \emph{``Rapports de Probabilites de Transition E0 et E2 Dans $^{178}$Hf"},
    Nucl. Phys. \textbf{A185} (1972) 321.
\bibitem[\href{http://www.nndc.bnl.gov/nsr/nsrlink.jsp?1973Ca10,B}{1973Ca10}]{1973Ca10}
    M.H. Cardoso, P.F.A. Goudsmit and J. Konijn,
    \emph{``The Decay of $^{172}$Ta"},
     Nucl. Phys. \textbf{A205} (1973) 121.
\bibitem[\href{http://www.nndc.bnl.gov/nsr/nsrlink.jsp?1973VaYZ,B}{1973VaYZ}]{1973VaYZ}
    J. Van Klinken, \emph{et al.},
   \emph{``Systematic experiments on neutron deficient nuclei in the region $50 \leq (N,Z) \leq 82$"},
    (1973) KFK-1768.
\bibitem[\href{http://www.nndc.bnl.gov/nsr/nsrlink.jsp?1974Ca28,B}{1974Ca28}]{1974Ca28}
    M. Cailliau, \emph{et al.},
    \emph{``Un Noyau Loin de la Stabilite: Le $^{182}$Pt"},
     Journal De Physique \textbf{35} (1974) L233.
\bibitem[\href{http://www.nndc.bnl.gov/nsr/nsrlink.jsp?1974Dr02,B}{1974Dr02}]{1974Dr02}
     J.E. Draper, \emph{et al.},
     \emph{``In--beam measurement of E0 matrix elements in $^{72}$Se and $^{72}$Ge"},
      Phys. Rev. \textbf{C9} (1974) 948.
\bibitem[\href{http://www.nndc.bnl.gov/nsr/nsrlink.jsp?1974Ha63,B}{1974Ha63}]{1974Ha63}
    J.H. Hamilton, \emph{et al.},
    \emph{``E0/E2 transition strengths and interpretations of $0^+$, $2^+$ states in $^{178}$Hf"},
    Phys. Rev. \textbf{C10} (1974) 2540.
\bibitem[\href{http://www.nndc.bnl.gov/nsr/nsrlink.jsp?1974Sc03,B}{1974Sc03}]{1974Sc03}
    K. Schreckenbach,
  \emph{``Konversionselektronen von $^{156}$Gd nach Neutroneneinfang"},
   Z. Naturforsch  \textbf{29a} (1974) 17.
\bibitem[\href{http://www.nndc.bnl.gov/nsr/nsrlink.jsp?1975Kh05,B}{1975Kh05}]{1975Kh05}
   T.A. Khan, \emph{et al.},
   \emph{``Study of the excited $0^{+}$ states through conversion electron spectroscopy of neutron rich Zr and Mo
   fission products"},
    Z. Phys \textbf{A275} (1975) 289.
\bibitem[\href{http://www.nndc.bnl.gov/nsr/nsrlink.jsp?1975Sc32,B}{1975Sc32}]{1975Sc32}
    U. Schneider and U. Hauser,
    \emph{``On E0 transitions in the decay of $^{152}$Eu-m (9.3 h)"},
    Z. Phys. \textbf{A273} (1975) 239.
\bibitem[\href{http://www.nndc.bnl.gov/nsr/nsrlink.jsp?1976Mu03,B}{1976Mu03}]{1976Mu03}
    L.G. Multhauf, K.G. Tirsell and R.A. Meyer,
    \emph{``Collective excitations in $^{246}$Cm and the decay of $^{246}$Am$^{m}$"},
     Phys. Rev. \textbf{C13} (1976)  771.
\bibitem[\href{http://www.nndc.bnl.gov/nsr/nsrlink.jsp?1977Be23,B}{1976Mu03}]{1977Be23}
    R. B\'eraud,  \emph{et al.},
    \emph{``Band crossing in $^{186}$Hg"},
    Nucl. Phys. \textbf{A284} (1977) 221.
\bibitem[\href{http://www.nndc.bnl.gov/nsr/nsrlink.jsp?1977Co21,B}{1977Co21}]{1977Co21}
    J.D.~Cole, \emph{et al.},
   \emph{ "Shape coexistence in $^{186}$Hg and the decay of $^{186}$Tl"},
   Phys. Rev. \textbf{C16} (1977) 2010.
\bibitem[\href{http://www.nndc.bnl.gov/nsr/nsrlink.jsp?1977Dr08,B}{1977Dr08}]{1977Dr08}
    J.E. Draper, R.J. McDonald and N.S.P. King,
    \emph{``In--beam study of $^{206}$Pb level scheme including conversion electrons $\ge$ 40 keV"},
    Phys. Rev. \textbf{C16} (1977) 1594.
\bibitem[\href{http://www.nndc.bnl.gov/nsr/nsrlink.jsp?1980Ah03,B}{1980Ah03}]{1980Ah03}
    I. Ahmad and R.K. Sjoblom,
    \emph{``Low--spin states of $^{250}$Cf populated in the electron capture decay of 2.22-h $^{250}$Es"},
     Phys. Rev. \textbf{C22} (1980) 1226.
\bibitem[\href{http://www.nndc.bnl.gov/nsr/nsrlink.jsp?1980Sc13,B}{1980Sc13}]{1980Sc13}
    F. Schussler, \emph{et al.},
    \emph{``Discovery of a very low-lying $0^+$ state in $^{98}$Sr and shape coexistence implication in $^{98}$Sr"},
    Nucl. Phys. \textbf{A339} (1980)  415.
\bibitem[\href{http://www.nndc.bnl.gov/nsr/nsrlink.jsp?1981Pa10,B}{1981Pa10}]{1981Pa10}
    A. Passoja,\emph{et al.},
    \emph{``High-Resolution Study of E0 Internal Pair Decay of Excited 0$^{+}$ States in $^{58,60,62}$Ni."},
     Nucl. Phys. \textbf{A363} (1981) 399.
\bibitem[1980Pa-thesis]{1980Pa_thesis}
    A. Passoja,
    PhD Thesis,
    Univ. of Jyv\"askyl\"a, Finland (1980)
\bibitem[\href{http://www.nndc.bnl.gov/nsr/nsrlink.jsp?1982Go02,B}{1982Go02}]{1982Go02}
    U. Goerlach, \emph{et al.},
    \emph{``Lowest $\beta$--vibrational phonon in the second minima of $^{236,238}$U"},
     Phys. Rev. Lett. \textbf{48} (1982)  1160.
\bibitem[\href{http://www.nndc.bnl.gov/nsr/nsrlink.jsp?1984Co17,B}{1984Co17}]{1984Co17}
    J.D.~Cole \emph{et al.},
    \emph{``Decay of $^{188}Tl$ and observed shape coexistence in the bands of $^{188}$Hg"},
     Phys. Rev. \textbf{C30} (1984) 1267.
\bibitem[\href{http://www.nndc.bnl.gov/nsr/nsrlink.jsp?1984Ho02,B}{1984Ho02}]{1984Ho02}
    R.W. Hoff, \emph{et al.},
    \emph{``Levels of $^{244}$Cm populated by the beta decay of 10-h $^{244}$Am$^{g}$
     and 26-min $^{244}$Am$^{m}$"},
      Phys. Rev. \textbf{C29} (1984) 618.
\bibitem[\href{http://www.nndc.bnl.gov/nsr/nsrlink.jsp?1985Pa07,B}{1985Pa07}]{1985Pa07}
     A. Passoja, \emph{et al.},
     \emph{``Electromagnetic Decay of Excited $0^+$ States in $^{64,66,68}Zn$."},
      Nucl. Phys. \textbf{A438} (1985) 413.
\bibitem[\href{http://www.nndc.bnl.gov/nsr/nsrlink.jsp?1985Pa15,B}{1985Pa15}]{1985Pa15}
     A. Passoja, \emph{et al.},
      \emph{``E0 Transitions in $^{70}Ge$ and Shape-Coexistence Interpretation of Even-Mass Ge Isotopes"},
       Nucl. Phys. \textbf{A441} (1985) 261.	

\bibitem[\href{http://www.nndc.bnl.gov/nsr/nsrlink.jsp?1985Ra21,B}{1985Ra21}]{1985Ra21}
    V.~Rahkonen and T.~Lonnroth,
    \emph{``Low- and medium-spin in the N = 124 isotones $^{208}$Po and $^{209}$At"},
    Z. Phys. \textbf{A322} (1985) 333.
\bibitem[\href{http://www.nndc.bnl.gov/nsr/nsrlink.jsp?1986Ka07,B}{1986Ka07}]{1986Ka07}
    J. Kantele, \emph{et al.},
    \emph{``E0 transitions in $^{202,204}$Pb and intruder--state systamatics of even-even lead isotopes"},
    Phys. Lett. \textbf{B171} (1986) 151-154.
\bibitem[\href{http://www.nndc.bnl.gov/nsr/nsrlink.jsp?1986Pa19,B}{1986Pa19}]{1986Pa19}
    A.~Passoja,
    \emph{``First measurement of K--shell--electron to internal-pair conversion probability ratios for E0
     transitions in (fp) shell region"},
    Z. Phys. \textbf{A325} (1986) 299.
\bibitem[\href{http://www.nndc.bnl.gov/nsr/nsrlink.jsp?1986Su11,B}{1986Su11}]{1986Su11}
    A.R.H. Subber, \emph{et al.},
    \emph{``E0 transitions in the $\gamma$-unstable nucleus $^{124}$Te"},
    J. Phys. (London) \textbf{G12} (1986) 881.
\bibitem[\href{http://www.nndc.bnl.gov/nsr/nsrlink.jsp?1987Fa07,B}{1987Fa07}]{1987Fa07}
    K.~Farzin, \emph{et al.},
   \emph{``Conversion coefficients and E0 transitions in $^{102,104,106}$Pd"},
    Z. Phys. \textbf{A326}   (1987) 401.
\bibitem[\href{http://www.nndc.bnl.gov/nsr/nsrlink.jsp?1987Ju06,B}{1987Ju06}]{1987Ju06}
    R. Julin, \emph{et al.},
    \emph{``E0 study of 0$^{+}$ states near 5 MeV in $^{208}$Pb"},
    Phys. Rev. \textbf{C36} (1987) 1129.
\bibitem[\href{http://www.nndc.bnl.gov/nsr/nsrlink.jsp?1987Sa41,B}{1987Sa41}]{1987Sa41}
    M. Sakai, \emph{et al.},
    \emph{``Anomalous electron intensity ratio of the E0 internal conversion in the transitions of $^{154}$Gd"},
    Nucl. Phys. \textbf{A473} (1987) 317.
\bibitem[\href{http://www.nndc.bnl.gov/nsr/nsrlink.jsp?1987Va09,B}{1987Va09}]{1987Va09}
    P. Van Duppen, \emph{et al.},
    \emph{``$\beta ^{+}$/electron-capture decay of $^{192,194,196,198,200}$Bi: Experimental evidence for low lying 0$^{+}$ states"},
    Phys. Rev. \textbf{C35} (1987) 1861.
\bibitem[\href{http://www.nndc.bnl.gov/nsr/nsrlink.jsp?1987Ya13,B}{1987Ya13}]{1987Ya13}
    S.W. Yates, \emph{et al.},
    \emph{``E0 Decay of $0^+$ States in $^{146}$Gd: search for two-phonon octupole excitations"},
    Phys. Rev. \textbf{C36} (1987) 2143.
\bibitem[\href{http://www.nndc.bnl.gov/nsr/nsrlink.jsp?1990Ad07,B}{1990Ad07}]{1990Ad07}
    I. Adam, \emph{et al.},
    \emph{``The decay of $^{164}$Tm $\rightarrow$ $^{164}$Er"},
    Bull. Acad. Sci. USSR, Phys. Ser. \textbf{54}(9) (1990) 134.
\bibitem[\href{http://www.nndc.bnl.gov/nsr/nsrlink.jsp?1990Tr01,B}{1990Tr01}]{1990Tr01}
    W.H. Trzaska, \emph{et al.},
    \emph{``Comparison of experimental and calculated K/L ratios of E0 transitions in some heavy nuclei"},
    Z. Phys. \textbf{A335} (1990) 475.
\bibitem[\href{http://www.nndc.bnl.gov/nsr/nsrlink.jsp?1991Ch05,B}{1991Ch05}]{1991Ch05}
    B. Chand, \emph{et al.}
    \emph{``Study of the radioactive decays of $^{140}$Ba and $^{140}$La"},
    Can. J. of Phys. \textbf{69} (1991) 90.
\bibitem[\href{http://www.nndc.bnl.gov/nsr/nsrlink.jsp?1991Ma07,B}{1991Ma07}]{1991Ma07}
    P.F. Mantica, Jr., \emph{et al.},
    \emph{``E0 transitions and $0^+$ levels in $^{136}$Xe"},
    Phys. Rev. \textbf{C43} (1991) 1696.
\bibitem[\href{http://www.nndc.bnl.gov/nsr/nsrlink.jsp?1994Ki01,B}{1994Ki01}]{1994Ki01}
    T.~Kib\'{e}di, \emph{et al.},
    \emph{``Low--spin non--yrast states and collective excitations in $^{174}$Os, $^{176}$Os, $^{178}$Os,
    $^{180}$Os, $^{182}$Os and $^{184}$Os"},
    Nucl. Phys. \textbf{A567} (1994) 183.
\bibitem[\href{http://www.nndc.bnl.gov/nsr/nsrlink.jsp?1999Da18,B}{1999Da18}]{1999Da18}
    P.M.~Davidson, \emph{et al.},
    \emph{``Non--yrast states and shape co-existence in light Pt isotopes"},
    Nucl. Phys. \textbf{A657} (1999) 219.
\bibitem[\href{http://www.nndc.bnl.gov/nsr/nsrlink.jsp?1999Le61,B}{1999Le61}]{1999Le61}
    Y. Le Coz, \emph{et al.},
    \emph{``Evidence of multiple shape-coexistence in $^{188}$Pb"},
     EPJdirect \textbf{A3} (1999) 1.

\bibitem[\href{http://www.nndc.bnl.gov/nsr/nsrlink.jsp?2001Ki10,B}{2001Ki10}]{2001Ki10}
    T.~Kib\'{e}di, \emph{et al.},
    \emph{``Low--spin non--yrast states in light tungsten isotopes and the evolution of shape coexistence"},
    Nucl. Phys. \textbf{A688} (2001) 669.
\bibitem[\href{http://www.nndc.bnl.gov/nsr/nsrlink.jsp?2014Su05,B}{2014Su05}]{2014Su05}
    S. Suchyta, \emph{et al.}, 
    \emph{``Shape coexistence in $^{68}$Ni"},
    Phys.Rev. \textbf{C89}  (2014) 021301.
\bibitem[2018ErXX]{2018Eriksen}
    T.K. Eriksen, \emph{``Investigation of the Hoyle state in $^{12}$C and the related triple alpha reaction rate"},
    Ph.D., Australian National University, 2018
\bibitem[\href{https://www.nndc.bnl.gov/nsr/nsrlink.jsp?1971Be76,B}{1971Be76}]{1971Be76}
    B.N.~Belyaev, S.S.~Vasilenko, D.M.~Kaminker,
   \emph{``Pair Conversion in $^{42}$Ca"},
     Bull. Acad. Sci. USSR, Phys. Ser. \textbf{35}  (1972) 742.

\end{theDTbibliography}

\end{document}